\documentclass[journal,12pt,onecolumn,draftclsnofoot]{IEEEtran}

\usepackage{amsthm,amsmath}
\usepackage[utf8]{inputenc}

\usepackage{hyperref}
\usepackage{dsfont} 
\usepackage{amsfonts}
\usepackage{amssymb}
\usepackage{mathrsfs}
\usepackage{amsmath}
\usepackage{algorithm}
\usepackage{algpseudocode}
\usepackage{setspace}
\usepackage{algorithmicx}
\usepackage{algcompatible}
\usepackage{graphicx}
\usepackage{epsfig}
\usepackage{epstopdf}
\usepackage{subfigure}

\theoremstyle{plain} 
\newtheorem{theorem}{Theorem}[]

\newtheorem{proposition}{Proposition}[]

\DeclareMathOperator{\Tr}{Tr}

\IEEEoverridecommandlockouts

\begin{document}

\title{User Grouping and Resource Allocation in Multiuser MIMO Systems under SWIPT}

\author{Javier Rubio, Antonio Pascual-Iserte\\
Dpt. of Signal Theory and Communications\\Universitat Politècnica de Catalunya (UPC)\\ C. Jordi Girona 1-3, Building D5, 08034 Barcelona, Spain\\antonio.pascual@upc.edu \thanks{Partial results of this paper have been presented previously by the same authors at conference IEEE GLOBECOM 2013 (@2013 IEEE: some materials are reprinted, with permission, from \cite{rub:13d}).\\DOI: 10.1186/s13638-019-1460-y}}

\markboth{Accepted paper at EURASIP Journal on Wireless Communications and Networking (2019) 2019:164}
{}

\maketitle

\begin{abstract}
This paper considers a broadcast multiple-input multiple-output (MIMO) network with multiple users and simultaneous wireless information and power transfer (SWIPT). In this scenario, it is assumed that some users are able to harvest power from radio frequency (RF) signals to recharge batteries through wireless power transfer from the transmitter, while others are served simultaneously with data transmission. The criterion driving the optimization and design of the system is based on the weighted sum rate for the users being served with data. At the same time, constraints stating minimum per-user harvested powers are included in the optimization problem. This paper derives the structure of the optimal transmit covariance matrices in the case where both types of users are present simultaneously in the network, particularizing the results to the cases where either only harvesting nodes or only information users are to be served. The tradeoff between the achieved weighted sum rate and the powers harvested by the user terminals is analyzed and evaluated using the rate-power (R-P) region. Finally, we propose a two-stage user grouping mechanism that decides which users should be scheduled to receive information and which users should be configured to harvest energy from the RF signals in each particular scheduling period, this being one of the main contributions of this paper. 
\end{abstract}


\section{Introduction}
Currently, one of the main limiting factors of user terminals is the very limited lifetime of their batteries. One of the solutions to enhance this lifetime is based on energy harvesting technology, by means of which terminals can collect ambient energy without being physically plugged in \cite{paradiso:05}, \cite{sudevalayam:11}. This is especially important in scenarios where the nodes are located in places where the replacement or recharge of batteries is very difficult, costly, or even impossible (e.g., wireless sensor networks). However, this is not the only scenario that can benefit from energy harvesting technology. For example, in cellular communications, the number of users has increased exponentially, together with the rates of the communications, but the battery lifetimes are very short. In this case, energy harvesting could play a beneficial role.

Wind and solar energy compose the classical and best known examples of sources of energy harvesting, although other technologies could also be considered, such as those applied to moving sensors (this may be the case for cellular phones) based on piezoelectric technologies. In recent years, there have also been significant advances in the use of radio frequency (RF) signals as a source of energy scavenging. Although initial experimental measurements showed that the actual strengths of the received electric fields were significant only when the distances between the transmitters and the receivers are rather short \cite{paradiso:05}, current technological developments (both in terms of harvesting hardware and system features) allow for effectively taking advantage of RF energy harvesting in new scenarios \cite{zhang:19}. In fact, this is a trend that is being adopted in the design of current and future networks based on short distances (e.g., femtocells \cite{chandrasekhar:08}). Due to this, users will be able to be served with the higher bit rates that newer applications require. These low distances will allow for mobile terminals to be able to harvest power from the received radio signals when they are not detecting information data. This is commonly termed \emph{wireless power transfer} (see \cite{lu:15} for an extensive review of this technique) and is one of the main topics of this paper.

\subsection{Related Work}
The first work that introduced the concept of simultaneous wireless information and power transfer (SWIPT) was \cite{varshney:08}. In that work, it was proven, for the single-antenna additive white Gaussian noise (AWGN) channel, that the data rate and power transfer are related in a nontrivial way. The extension of the previous conclusion to the frequency-selective single-antenna AWGN channels was addressed later in \cite{grover:10}. Much effort has been put forward lately to come up with beamforming design strategies for the SWIPT framework. In \cite{zhang:13b}, the authors considered a multiple-input multiple-output (MIMO) system. In that paper, it was assumed that the transmitter was able to simultaneously transmit data and power to a single receiver. Two receiver architectures were considered able to combine both information and power sources simultaneously. In \cite{park:13} and \cite{park:14}, the authors considered an MIMO network consisting of multiple transmitter-receiver pairs with co-channel interference. The study in \cite{park:13} focused on the case with two transmitter-receiver pairs, whereas in \cite{park:14}, the authors generalized \cite{park:13} by considering that $k$ transmitter-receiver pairs were present. In \cite{zong:16}, the authors considered an MIMO system with single-stream transmission. In contrast to previous works, where the system rate was optimized, the objective of the above authors was to minimize the overall power consumption with minimum signal to interference and noise ratio (SINR) constraints and per-user harvesting constraints. Multiuser broadcast networks can also be found under the framework of multiple-input single-output (MISO) beamforming, as in \cite{xu:14} and \cite{shi:14}. The main difference between our work and previous works is that we assume a broadcast multiuser multistream MIMO network, which has not been considered before.

Although, in this paper, we assume that the channel state information (CSI) is known at the transmitter, there are some works that can be referenced in which techniques for optimizing the training under the SWIPT framework are presented \cite{zeng:15}, \cite{xu:16}. In particular, \cite{zeng:15} studies the design of an efficient channel acquisition method for a point-to-point MIMO SWIPT system by exploiting the channel reciprocity. Additionally, a worst-case robust beamforming design was proposed in \cite{xiang:12}, in which imperfect CSI at the transmitter was assumed. Another strategy is to overcome this CSI feedback, as was done with implicit beamforming in \cite{shepard:12}.

In this paper, we propose some user grouping techniques in which, from frame to frame, it is decided which users will receive information data and which users will harvest energy from RF signals. There are some works in the literature that deal with user scheduling in the SWIPT framework, but they consider a single-input single-output (SISO) system. Therefore, the scheduling presented in those papers is purely the temporal scheduling of users. Among those works, \cite{zhang:13a} introduced time scheduling between information and energy transfer and derived the optimal switching policy considering time-varying co-channel interference. The receiver therefore replenished the battery opportunistically via wireless power transfer from the unintended interference and/or the intended signal sent by the transmitter. Then, in \cite{morsi:15}, the authors studied downlink multiuser scheduling for a time-slotted system with SWIPT. In particular, in each time slot, a single user is scheduled to receive information, whereas the remaining users opportunistically harvest energy from ambient signals. Finally, in \cite{ding:14}, the authors considered a multiuser cooperative network, where $M$ source-destination SISO pairs communicate with each other via a relay with energy harvesting capabilities. The key idea is to select a subset of those $M$ pairs to communicate through the relay. In contrast to those works, in this paper, we present a spatial user grouping strategy since a multiuser MIMO system is considered, and multiple users therefore can be served simultaneously at each scheduling period. We also implement temporal scheduling, as those spatial user groups change over time due to the dynamics of the batteries and the historic user performance. 

Finally, we want to mention that there are also several works in the literature dealing with user grouping strategies in the multiple-antenna scenario, although none of them has considered the general case addressed in this paper, that is, the problem of grouping and scheduling users in a limited-energy system with SWIPT, a multi-antenna transmitter and multiple multi-antenna receivers, and taking into account the temporal evolution of the states of the batteries. For example, in \cite{shen:15}, a grouping strategy is developed for the case of a multiuser system, with one multi-antenna transmitter and single-antenna receivers (instead of multi-antenna receivers, as we consider in our paper) based on zero-forcing (ZF) precoding but without considering power transfer and without including the effect of the batteries. To the best of the authors’ knowledge, the most recent paper related to our work is \cite{dau:19}. That paper addresses the same setup as \cite{shen:15}, that is, one multi-antenna transmitter and single-antenna receivers, where the transmitter is enabled with hybrid precoding and the digital beamformers are designed according to the ZF criterion. The paper designs the transmitter by simultaneously considering the transmission of data and power through harvesting power splitting. Due to the complexity of the problem, \cite{dau:19} decouples the design of the user grouping (that is based on the correlation of the equivalent channels), the beamformers, and the power/harvesting parameters. The design of the power allocation and the power splitting parameters is addressed through an optimization problem, aiming at maximizing the sum rate while requiring minimum rates and harvested powers. Our paper generalizes the work of \cite{dau:19} by considering multiple antennas at the receivers and by not decoupling the design into several substages, which is a suboptimum approach. In this sense, we include the design of the beamformers into the optimization problem, improve the user grouping by considering the result of the optimization problem beyond the channel correlation, and explicitly take into account the states of the batteries and their time evolution in the grouping strategy. For these reasons, the techniques presented in the previous papers cannot be compared with ours due to the fact that they only consider single-antenna receivers and do not include the states of the batteries. There are more papers in the literature, but they consider even more simplified system assumptions than the previous two \cite{shen:15} and \cite{dau:19} and, therefore, are not cited here for the sake of brevity.

\subsection{Contributions}
In this paper, we extend the previous works by addressing a multiuser multistream MIMO system, where multiple information and energy harvesting receivers are present and where we explicitly consider other power consumption sources in the system design. The receivers are considered constrained by the system's battery dynamics, and in this sense, the batteries need to be recharged to increase their lifetimes. In the multiuser MIMO SWIPT framework, there are two groups of users to be served: one for power reception to recharge the batteries, and the other for information reception. Thus far in the literature of MIMO beamforming techniques, authors have considered that these two sets of users were predefined and fixed. In this paper, we propose some user grouping techniques that may change frame to frame to maximize the system throughput and/or fairness among users. Additionally, only single-stream communications have been considered for the broadcast scenario so far. The problem of maximizing the multistream sum rate for the multiuser MIMO scenario is very difficult and nonconvex \cite{goldsmith:03}. For this reason, we propose the use of a conventional block-diagonalization (BD) \cite{spencer:04} simplification used extensively in the literature \cite{zhang:10a} and generalize most of the works found in the literature by considering multistream communications. 

The alternative, that is, not forcing BD and allowing for the presence of interference, results in a nonconvex highly complex problem that we have addressed in our recent journal paper \cite{rub:16a}. The complexity of that problem is such that the whole paper is dedicated exclusively to the proposal of numerical algorithms to find a local optimum of the nonconvex problem. In that paper, we assume that the user grouping is fixed and known, and we do not consider the design of those user groups, the performance evaluation of the temporal behavior of the system, the presence of any scheduler, or the presence of user batteries.

Compared to the works presented in the previous section, the main contributions of our work can be highlighted as follows:
\begin{itemize}
	\item We consider a multiuser multistream MIMO broadcast transmission strategy in which both the transmitter and receivers are provided with multiple antennas. The system weighted sum rate with individual per-user harvesting constraints is considered in the proposed transmission strategy design. We also take into account the state of the batteries of the terminals in the proposed strategy. We study particular cases in which only information users and only harvesting users are present in the system.
	\item We develop an efficient algorithm that computes the optimal precoding matrices for the multiuser MIMO broadcast network setup mentioned previously.
	\item The fundamental (multidimensional) tradeoff between system performance and (per-user) harvested energy is studied and characterized, placing emphasis on and giving specific closed-form expressions for some particular cases of interest. 
	\item We incorporate power consumption models at the transmitter and receivers. In particular, we consider the decoding power consumption at the receivers and its impact on system performance. 
	\item Finally, we develop harvesting-constrained user grouping schemes that employ a two-stage user scheduling mechanism that runs at different time scales. In the first stage, a subset of users are grouped to be candidates for information reception, and a subset of users are grouped to be candidates for harvesting users. Out of these selected users, in the second stage, we perform the final user information and harvesting grouping, with the aim of enhancing the system throughput and/or fairness among users. 
\end{itemize}

The work developed in this paper extends our previous work presented in a conference paper \cite{rub:13d}. The main differences and new contributions with respect to that conference version are summarized as follows. First, in this journal version, we have assumed that the system evolves over time, and we therefore have considered a generalized formulation and the inclusion of some power consumption sinks that affect the battery dynamics. Second, we have assumed that user groups are not fixed and known by the transmitter; hence, user grouping strategies have been derived, resulting from the consideration of an optimization of the system performance over time. Finally, we have included a full simulation section that evaluates the system performance over time.

\subsection{Organization of the Paper}
The remainder of this paper is organized as follows. In Section \ref{sec_sysm}, we present the system model. In Section \ref{sec_joint}, we present the formulation of the most general user grouping and resource allocation strategy. We formulate and justify the simplifications that we consider in this paper to solve such a complex problem. Section \ref{sr-ie} covers the precoder design for simultaneous data and power transfer. We also address the characterization of the fundamental tradeoff between data and power transfer. In Section \ref{sec_sched}, we present a scheduling mechanism to decide which users should be scheduled in each particular user set. The overall algorithm including all the stages, that is, the user grouping and the resource allocation, is described in detail in Section \ref{sec_overall_algorithm}. Section \ref{num_sec} presents some numerical results of the proposed techniques. Finally, conclusions are drawn in Section \ref{sec_con}.

\subsection{Notation Used in the Paper}
The notation that will be used in this paper is detailed in Table \ref{table:notation}.

\begin{table}[t]
	\renewcommand{\arraystretch}{0.98}
	\caption{\textsc{Notation used in the paper.}}
	\label{table:notation}
	\begin{center}
		\begin{tabular}{| c |  l | }
			\hline
			$\mathcal{A}$ & set \\ \hline
			$\mathcal{A}=\{a_1,a_2,\dots\}$ & set $\mathcal{A}$ containing the elements $\{a_1,a_2,\dots\}$ \\ \hline
			$|\mathcal{A}|$ & number of elements in set $\mathcal{A}$ \\ \hline
			$a \in \mathcal{A}$ & $a$ belongs to set $\mathcal{A}$ \\ \hline
			$\mathcal{A} \setminus a$ & set resulting from subtracting $a$ from set $\mathcal{A}$ \\ \hline
			$\emptyset$ & empty set \\ \hline
			$\mathcal{A} \subseteq \mathcal{B}$ & set $\mathcal{A}$ is included in or equal to set $\mathcal{B}$ \\ \hline
			$\mathcal{A}\cap\mathcal{B}, \mathcal{A}\cup\mathcal{B}$ & intersection of sets $\mathcal{A}$ and $\mathcal{B}$, union of sets $\mathcal{A}$ and $\mathcal{B}$ \\ \hline
			$\mathbf{a}, \mathbf{A}$ & vector $\mathbf{a}$, matrix $\mathbf{A}$ \\ \hline
			$\mathbf{a}^T, \mathbf{A}^T$ & transpose of vector $\mathbf{a}$, matrix $\mathbf{A}$ \\ \hline
			$\mathbf{a}^H, \mathbf{A}^H$ & Hermitian (transpose conjugated) of vector $\mathbf{a}$, matrix $\mathbf{A}$ \\ \hline
			$\mbox{Tr}(\mathbf{A}), \det(\mathbf{A})$ & trace of matrix $\mathbf{A}$, determinant of matrix $\mathbf{A}$ \\ \hline
			$\mathbf{A}\succeq 0$ & matrix $\mathbf{A}$ is positive semidefinite \\ \hline
			$||\mathbf{a}||$ & norm-2 of vector $\mathbf{a}$ \\ \hline
			$\mathbb{C}^{m\times n}$ & set of complex matrices of size $m\times n$ \\ \hline
			$\mathbf{I}_n$ & identity matrix of size $n\times n$ \\ \hline
			$\mathbb{E}[\cdot]$ & expectation \\ \hline	
			$=, \triangleq ,\neq$ & equal, equal by definition, different \\ \hline
			$>, \geq, <, \leq$ & higher, higher or equal, lower, lower or equal \\ \hline
			$\log(\cdot), \exp(\cdot)=\mbox{e}^{(\cdot)}$ & logarithm, exponential \\ \hline
			$n!$ & factorial of $n$ \\ \hline
			$\sum$ & summation \\ \hline
			$\min, \max$ & minimum, maximum \\ \hline
			$(x)^b_a$ & $(x)^b_a = \min\{\max\{a,x\},b\}$ \\ \hline
			$a^b$ & $a$ to $b$ \\ \hline
			$\forall$ & for all \\ \hline
			$\mathop{\text{maximize}}_{x_1,x_2,\dots}$ & maximization with respect to variables $x_1,x_2,\dots$ \\ \hline
			$\mathop{\text{minimize}}_{x_1,x_2,\dots}$ & minimization with respect to variables $x_1,x_2,\dots$ \\ \hline
			$x^\star$ & optimum value of $x$ \\ \hline
			$f^{-1}(\cdot)$ & inverse function \\ \hline
			$x \leftarrow y$ & $x$ is updated with $y$\\
			\hline
		\end{tabular}
	\end{center}
\end{table}

\section{System Model}
\label{sec_sysm}
\subsection{Signal Model}
\begin{figure}[h!]
	\centering
	\textbf{Broadcast multiuser system}\\
	\includegraphics[width=0.9\textwidth]{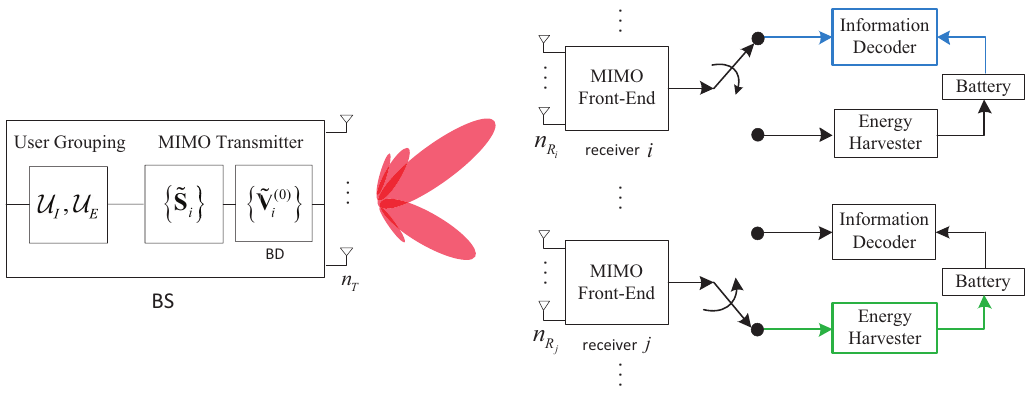}
	\caption{Schematic representation of the downlink broadcast multiuser communication system. Note that each user can switch from an information decoder receiver to an energy harvester receiver. This switching can be implemented technologically, as mentioned in papers such as \cite{hua:16}.}
	\label{fig_arc}
\end{figure}

We consider a wireless broadcast system consisting of one base station (BS) transmitter equipped with $n_T$ antennas and a set of $K$ receivers, denoted as $\mathcal{U}_T = \{1, 2, \dots , K\}$, where the $k$-th receiver is equipped with $n_{R_k}$ antennas, as depicted in Fig. \ref{fig_arc}. 

We index frames by $t \in \mathcal{T} \triangleq \{1,\dots,T\}$ with a duration of $T_f$ seconds each. We assume block fading channels, that is, the channels remain constant within a frame but change from frame to frame. The equivalent baseband channel from the BS to the $k$-th receiver is denoted by $\textbf{H}_k(t) \in \mathbb{C}^{n_{R_k}\times n_{T}}$. It is also assumed that the set of matrices $\{\textbf{H}_k(t)\}$ is known to the BS and to the corresponding receivers. The case of imperfect CSI is beyond the scope of the paper.

The set of users is partitioned into two subsets, as mentioned in the introduction. One of the sets contains the users that receive information, denoted as $\mathcal{U}_I(t) \subseteq \mathcal{U}_T$ and $|\mathcal{U}_I(t)| = N$, and the other set, $\mathcal{U}_E(t) \subseteq \mathcal{U}_T$, $|\mathcal{U}_E(t)| = M$, contains the users that harvest energy from the power radiated by the BS, which is used to transmit signals to the information receivers. Note that the previous sets depend on $t$, as the specific users in each of them may change from frame to frame. The numbers of users in each set, $N$ and $M$, may change from frame to frame as well, as will be explained later in the paper. We assume that a given user is not able to simultaneously decode information and harvest energy. This forces a user to either receive information or harvest energy during the whole frame, i.e., during the scheduling period, which is a reasonable choice if the scheduling periods are short. That translates into disjoint subsets, i.e., $\mathcal{U}_I(t) \,\cap\, \mathcal{U}_E(t) = \emptyset$,  $|\mathcal{U}_I(t)| + |\mathcal{U}_E(t)| \le K$.\footnote{Let us assume for the moment that not all users must be in any group. As will be shown later, some of the users may not be selected for any group in a given scheduling period.} To simplify the notation when needed, we will assume that the indexing of users is such that $\mathcal{U}_I(t) = \{1,2,\dots,N\}$ and $\mathcal{U}_E(t) = \{N+1,N+2,\dots,N+M\}$.\footnote{At the beginning of each frame, once the groups have been decided, the users are indexed again in such a way that the first $N$ users are information users and the following $M$ users are harvesting users.} We will assume that $n_T > n_R - \min_k\{n_{R_k}\}$ is fulfilled, being $n_R = \sum_{k\in\mathcal{U}_I}n_{R_k}$.\footnote{This assumption corresponds to a necessary constraint to be applied when block diagonalization (BD) is used \cite{spencer:04}, as will be explained in more detail in Section \ref{sr-ie}.}

As far as the signal model is concerned, the received signal for the $i$-th information receiver at the $n$-th time instant within the $t$-th frame can be modeled as
\begin{eqnarray}
\hspace{-7mm}\textbf{y}_i(n,t) = \textbf{H}_i(t)\textbf{B}_i(t)\textbf{x}_i(n,t) + \textbf{H}_i(t)\sum_{\substack{k\in\mathcal{U}_I(t)\\k\neq i}}\textbf{B}_k(t)\textbf{x}_k(n,t) + \textbf{w}_i(n,t) \in \mathbb{C}^{n_{R_i}\times 1}, & & \label{sig_model} \\ \forall i \in \mathcal{U}_I(t).& &
\nonumber
\end{eqnarray}
In the previous notation, $\textbf{B}_i(t)\textbf{x}_i(n,t)$ represents the transmitted signal for user $i \in \mathcal{U}_I(t)$, where $\textbf{B}_i(t)\in \mathbb{C}^{n_{T}\times n_{S_i}}$ is the precoder matrix, and $\textbf{x}_i(t)\in \mathbb{C}^{n_{S_i}\times 1}$ represents the information symbol vector. $n_{S_i}$ denotes the number of streams assigned to user $i \in \mathcal{U}_I(t)$, and we assume that $n_{S_i} = \min\{n_{R_i}, n_T – (n_R – n_{R_i})\} \, \forall i \in \mathcal{U}_I(t)$ is fulfilled\footnote{In fact $\min\{n_{R_i}, n_T-(n_R-n_{R_i})\}$ is an upper bound for the actual number of active streams. Such a number will be obtained from the solution of the corresponding optimization problems presented in this paper (in Section \ref{sr-ie}).}. The transmit covariance matrix is $\textbf{S}_i(t) = \textbf{B}_i(t)\textbf{B}_i^H(t)$ if we assume, without loss of generality (w.l.o.g.), that $\mathbb{E}\left[\textbf{x}_i(n,t)\textbf{x}_i^H(n,t)\right] = \textbf{I}_{n_{S_i}}$. $\textbf{w}_i(n,t)\in \mathbb{C}^{n_{R_i}\times 1}$ denotes the receiver noise vector, which is considered white and Gaussian with $\mathbb{E}\left[\textbf{w}_i(n,t)\textbf{w}_i^H(n,t)\right] = \textbf{I}_{n_{R_i}}$\footnote{We assume that noise power $\sigma^2 = 1$ w.l.o.g.; otherwise, we could simply apply a scale factor at the receiver and rescale the channels accordingly.}. Note that the middle term of \eqref{sig_model} is an interference term usually known as multiuser interference (MUI).

Let $\tilde{\textbf{x}}(n,t) = \textbf{B}(t)\textbf{x}(n,t)$ denote the signal vector transmitted by the BS, where the joint precoding matrix is defined as $\textbf{B}(t) = [\textbf{B}_1(t) ,\dots , \textbf{B}_N(t)] \in \mathbb{C}^{n_T \times n_S}$, where $n_S = \sum_{i=1}^Nn_{S_i}$ is the total number of streams of all information users, and the data vector is $\textbf{x}(n,t) = \left[\textbf{x}_1^T(n,t) ,\dots , \textbf{x}_N^T(n,t)\right]^T \in \mathbb{C}^{n_S \times 1}$. $\tilde{\textbf{x}}(n,t)$ must satisfy the power constraint formulated as $\mathbb{E}[\|\tilde{\textbf{x}}(n,t)\|^2] = \sum_{i=1}^N\Tr(\textbf{S}_i(t))\le P_{T}$, where $P_{T}$ represents the total radiated power at the BS, assuming that the information symbols of different users are independent and zero-mean.

Let us model the total power harvested by the $j$-th user during the $t$-th frame, denoted by $\bar{Q}_j(t)$, from all receiving antennas to be proportional to that of the equivalent baseband signal, i.e.,
\begin{equation}
\bar{Q}_j(t) =\zeta_j \sum_{i\in \mathcal{U}_I(t)}\mathbb{E}[\| \textbf{H}_j(t) \textbf{B}_i(t) \textbf{x}_i(n,t)\|^2], \quad \forall j \in \mathcal{U}_E(t),
\label{eqEn}\\
\end{equation}
where $\zeta_j$ is a constant that accounts for the loss in the energy transducer when converting the harvested power to electrical power to charge the battery. Note that, for simplicity, in \eqref{eqEn}, we have omitted the harvested power due to the noise term or other external RF sources since they can be assumed negligible. Based on this, \eqref{eqEn} can be written as
\begin{equation}
\bar{Q}_j(t) = \zeta_j \sum_{i\in\mathcal{U}_I(t)}\Tr(\textbf{H}_j(t)\textbf{S}_i(t)\textbf{H}_j^H(t)), \quad \forall j \in \mathcal{U}_E(t).
\label{eqEn2}
\end{equation}
For the sake of clarity, we will drop the time and frame dependence whenever possible.

\subsection{Power Consumption Models}
The energy consumed by the transceiver can be modeled as the energy consumed by the front-end plus the energy consumed by the coding/decoding stages (omitting for the moment the power radiated by the transmitter).\footnote{We consider a reference system, where the energy spent by the terminals is only driven by the power used for the communication (RF chains and decoding). It is true that we do not consider other sinks of energy consumption, such as the energy consumed by the application layer. In case we would want to include those, we could simply add the corresponding additional terms.} Although other works consider battery imperfections in their models \cite{bertrand:12}, we do not consider them in our work for the sake of simplicity. Note, however, that the strategy and formulation presented in this paper could be extended easily to incorporate those imperfections. In the following, we will comment briefly on the generic abstract approach followed in this paper to make the proposed strategies independent of the concrete model.
\begin{enumerate}
	\item \emph{Front-end Consumption:} as far as the transmitter is concerned, the components that consume energy are the high-power amplifier (HPA), the mixers, the filters, and other elements of the RF chain. Concerning the receiver, the front-end consumption usually depends on the condition on the channel, i.e., the signal to noise ratio (SNR) (in practice, the receiver should adapt the front-end according to the received power \cite{jensen:12}, an operation that requires some additional power). In the following, however, we assume that the component of the receiver front-end consumption that depends on the SNR is negligible, as it can be concluded from experimental measurements and is adopted in most works \cite{jensen:12}. We denote the energy consumed by the front-end at the transmitter and the receiver by $P_c^{t_x}$ and $P_c^{r_x}$, respectively.
	
	\item \emph{Coding/Decoding Consumption:} it is reasonable to consider the energy consumed by the coding stage at the transmitter negligible compared to the energy consumed by the front-end. This is illustrated and commented on in papers such as \cite{auer:11}. For this reason, we will not include coding consumption in our models. On the other hand, the decoding consumption must be included in the models since, as shown in \cite{grover:11}, \cite{ros:10}, such energy consumption is not negligible and can affect importantly the lifetime of the mobile terminal. There is a consensus about the fact that the decoding consumption increases with the data rate $R_i(t)$, $P_{\textrm{dec},i}(R_i(t))$. In \cite{rub:14a}, the authors presented different models for $P_{\textrm{dec},i}(R_i(t))$, but for the sake of generality, we will consider it a general function.
	
\end{enumerate}

Given the previous models, the total consumption at the transmitter (omitting for the moment the radiated power) only includes the front-end consumption as mentioned previously, and it therefore is denoted as
\begin{equation}
P_{\textrm{tot}}^{t_x} = P_{c}^{t_x}.
\end{equation}

On the other hand, the total power consumption at the $i$-th receiver is expressed as
\begin{equation}
P_{\textrm{tot},i}^{r_x}(R_i(t)) = P_{\textrm{dec},i}(R_i(t)) + P_{c}^{r_x}.
\label{Ptot}
\end{equation}
Note that the power consumption at the receiver is limited by the current battery level, which in the following will be denoted by $C_i(t)$ for user $i$. According to this, the data rate of a given information user (user $i$) during one frame must be constrained in order not to consume more energy when decoding than the current energy available at the battery $C_i(t)$. Hence,
\begin{equation}
T_f (P_{\textrm{dec},i}(R_i(t)) + P_{c}^{r_x}) \le C_i(t),
\end{equation}
which can be written in terms of a maximum rate constraint as 
\begin{equation}
R_i(t) \le R_{\max,i}(C_i(t)),  
\end{equation}
where $R_{\max,i}(C_i(t)) = P_{\textrm{dec},i}^{-1}\left(\frac{C_i(t)}{T_f}- P_{c}^{r_x}\right)$.

\subsection{Battery Dynamics}
We consider that each user terminal is provided with a finite battery capacity, the level of which decreases accordingly when the user receives and decodes data. The terminals are also able to recharge their batteries by means of collecting the power dynamically coming from the BS. 

The battery at the beginning of the $t$-th frame of the $i$-th information user served with a data rate $R_i(t-1)$ during the previous frame is denoted as
\begin{equation}
C_i(t) =  \left(C_i(t-1) - T_f P_{\textrm{tot},i}^{r_x}(R_i(t-1))\right)_0^{C_{\max}^i},\quad \forall i \in\mathcal{U}_I(t),
\end{equation}
where $(x)^b_a$ is the projection of $x$ onto the interval $[a, b]$, i.e., $(x)^b_a = \min\{\max\{a,x\},b\}$, $C_{\max}^i$ is the maximum battery level, and the function $P_{\textrm{tot,i}}^{r_x}(R_i(t-1))$ was defined in \eqref{Ptot}. Note that $C_i(t)$ has units of Joules. 

On the other hand, the battery at the beginning of the $t$-th frame of the $j$-th harvesting user is denoted as
\begin{equation}
C_j(t) = \left(C_j(t-1) + T_f\bar{Q}_j(t-1) - T_f P_{c}^{r_x}\right)_0^{C_{\max}^j},\quad \forall j \in\mathcal{U}_E(t),
\end{equation}
where $\bar{Q}_j(t-1)$ is the power harvested during the frame $t-1$.

The receivers must inform the BS about their battery level status to make decisions on whether to serve that user with information or with power. In this paper, we assume that the feedback channel is ideal and not rate-limited.

The power consumption and battery dynamics models, which are based on the state of the art and existing literature, were also used in a similar way by the same authors of this paper in their previous work \cite{rub:14a}.

\section{Joint Resource Allocation and User Grouping Formulation}
\label{sec_joint}
In this section, we formulate the joint design of the covariance matrices $\textbf{S}_i(t)$, the data rates $R_i(t)$, and the user grouping $\mathcal{U}_I(t), \,\mathcal{U}_E(t)$, based on the maximization of the weighted sum rate with individual power harvesting constraints for all time instants $t\in\mathcal{T}$. Given this, the problem is formulated through the following optimization problem (this formulation generalizes the problem defined in our previous conference paper \cite{rub:13d}):
\begin{alignat}{3}
\mathop{\text{maximize}}_{\substack{\{R_i(t),\, \textbf{S}_i(t)\}_{\forall i\in\mathcal{U}_I(t)}, \\\mathcal{U}_I(t),\, \mathcal{U}_E(t)}}& \quad \sum_{t\in\mathcal{T}}\sum_{i\in\mathcal{U}_I(t)} \omega_i(t)R_i(t) \label{op:wet_dp} \\
\textrm{subject to}\nonumber
\end{alignat}
\vspace{-1.5cm}
\begin{alignat}{2}
& C1:\sum_{i\in\mathcal{U}_I(t)}\Tr(\textbf{H}_j(t)\textbf{S}_i(t)\textbf{H}_j^H(t))  \ge Q_j, \hspace{-1cm}& \forall j \in \mathcal{U}_E(t),\,\forall t\in\mathcal{T} \nonumber \\
&  C2:\sum_{i\in\mathcal{U}_I(t)}\Tr(\textbf{S}_i(t)) + P_c^{tx}\le P_{\max},\hspace{-1cm}&\forall t\in\mathcal{T}\nonumber\\
&  C3:R_i(t)\le \log \det \left(\textbf{I} + \textbf{H}_i(t)\textbf{S}_i(t)\textbf{H}_i^H(t)\right), \hspace{-1cm}& \forall i \in \mathcal{U}_I(t),\,\forall t\in\mathcal{T}\nonumber\\
&  C4:R_i(t)\le R_{\max,i}(C_i(t)), \hspace{-1cm}& \forall i \in \mathcal{U}_I(t),\,\forall t\in\mathcal{T}\nonumber\\
&  C5: \textbf{H}_k(t)\textbf{S}_i(t)\textbf{H}_k^H(t) = 0, \hspace{-1cm}& \forall k\neq i, \,\,\, k, i \in \mathcal{U}_I(t),\,\forall t\in\mathcal{T} \nonumber\\
&  C6:  \textbf{S}_i(t) \succeq 0, \hspace{-1cm}& \forall i \in \mathcal{U}_I(t),\,\forall t\in\mathcal{T} \nonumber\\
&  C7:  C_i(t) =  \left(C_i(t-1) - T_f P_{\textrm{tot},i}^{r_x}(R_i(t-1))\right)_0^{C_{\max}^i},\hspace{-1cm}& \forall i \in\mathcal{U}_I(t),\,\forall t\in\mathcal{T}\nonumber\\
&  C8:  C_j(t) = \left(C_j(t-1) + T_f\bar{Q}_j(t-1) - T_f P_{c}^{r_x}\right)_0^{C_{\max}^j},\hspace{-1cm}& \forall j \in\mathcal{U}_E(t),\,\forall t\in\mathcal{T},\nonumber
\end{alignat}
where the weights $\omega_i(t)\ge0$ can be set to assign priorities to achieve fairness among the different users\footnote{A further discussion on how the weights $\omega_i(t)\ge0$ can be set to provide fairness will be introduced later in Section \ref{sec_sched}.}, $R_i(t) \le \log\det\left(\textbf{I} + \textbf{H}_i(t)\textbf{S}_i(t)\textbf{H}_i^H(t)\right)$ denotes the achievable data rate of the $i$-th user when considering linear precoding following a BD strategy \cite{spencer:04}, $Q_j = \frac{{\bar{Q}^{\min}}_j}{\zeta_j}$, where $\{\bar{Q}^{\min}_j\}$ is the set of minimum power harvesting constraints, and $P_{\max}$ is the available power at the BS. In fact, BD is applied through constraint $C5$, which forces the complete cancellation of the MUI, making the whole problem more tractable (as will be shown later in the paper). Notice that constraint $C1$ is associated with the minimum power to be harvested for a given user. In the case that another external energy harvesting source was available and the amount to be harvested could be estimated (or was fully known in advance), we could subtract such value from $Q_j$ accordingly. Constraint $C4$ assures that the information users do not spent more energy decoding the message than the current energy available at the battery. 

As we have already noted, we have assumed a linear precoding approach in the system formulation. Note that the optimum transmission policy in an MIMO broadcast channel is the well-known nonlinear dirty paper coding strategy \cite{goldsmith:03}. Nevertheless, that strategy has high computational demands and cannot be implemented in real time. Instead, much simpler linear transceiver designs have also been shown to achieve high capacities using much lower computational resources (see \cite{lee:06} for more details). Thus, for simplicity in the transmitter design, in this work, we force the precoder to be linear.

Two main difficulties arise when attempting to solve \eqref{op:wet_dp}. First, note that the solution for all time instants has to be found jointly. The reason is that resource allocation decisions at frame $t$ have an impact not only on that frame but also on future frames. Some researchers have attempted to solve harvesting (time-coupled) problems by assuming that the whole channel and harvesting realizations are known a priori, giving rise to offline approaches that are not implementable in real scenarios \cite{ho:12}, \cite{yang:12}. As we assume that only causal knowledge of the channel and the harvesting is available, we would have to resort to dynamic programming (DP) techniques \cite{bertsekas_dp} to find the optimal solution of problem \eqref{op:wet_dp}. However, these techniques usually require the implementation of extremely high complexity algorithms that are impractical in scenarios, where the set of variables to be optimized is large, and DP techniques therefore have been applied only in cases where the optimization variables are scalars \cite{blasco:13}, \cite{lopez_ramos:14}. The second difficulty that we find is that the user grouping must also be optimized jointly with the covariance matrices and the data rates. The user grouping variables are discrete, and the problem therefore becomes combinatorial. The optimum solution has to be found by applying some sort of combinatorial search among all possible user groups, increasing the overall complexity exponentially. 

Because we are interested in low-complexity solutions, we have to make some simplifications to problem \eqref{op:wet_dp} to make it more tractable, with the hope of finding a good suboptimum solution that is close to the global optimum solution of problem \eqref{op:wet_dp}. 

The first assumption that we consider is to decouple the problem in time and propose a separate per-frame optimization approach. With this approach, we solve the optimization problem at the beginning of each frame $t$, making decisions based on the current and past information on the battery levels. The optimization to solve is (we omit the time dependence for the sake of simplicity in the notation even though all these variables, including the information and harvesting users sets $\mathcal{U}_I$ and $\mathcal{U}_E$, change at each frame) as follows:
\begin{alignat}{3}
\mathop{\text{maximize}}_{\substack{\{R_i,\, \textbf{S}_i\}_{\forall i\in\mathcal{U}_I}, \\\mathcal{U}_I,\, \mathcal{U}_E}}& \quad \sum_{i\in\mathcal{U}_I} \omega_iR_i \label{op:wet_g} \\
\textrm{subject to}
&  \quad C1:\sum_{i\in\mathcal{U}_I}\Tr(\textbf{H}_j\textbf{S}_i\textbf{H}_j^H)  \ge Q_j, & \forall j \in \mathcal{U}_E \nonumber \\
&  \quad C2:\sum_{i\in\mathcal{U}_I}\Tr(\textbf{S}_i) + P_c^{tx}\le P_{\max}\nonumber\\
&  \quad C3:R_i\le \log \det \left(\textbf{I} + \textbf{H}_i\textbf{S}_i\textbf{H}_i^H\right), & \forall i \in \mathcal{U}_I\nonumber\\
&  \quad C4:R_i\le R_{\max,i}(C_i), & \forall i \in \mathcal{U}_I\nonumber\\
&  \quad C5: \textbf{H}_k\textbf{S}_i\textbf{H}_k^H = 0, & \forall k\neq i, \,\,\, k, i \in \mathcal{U}_I \nonumber\\
&  \quad C6:  \textbf{S}_i \succeq 0, & \forall i \in \mathcal{U}_I. \nonumber
\end{alignat}
Problem \eqref{op:wet_g}, which generalizes the one addressed in our previous paper \cite{rub:13d}, as weights are included to take into account the time evolution of the achieved rates, is still very difficult to solve, as it involves continuous and integer variables. Note that for a fixed set of groups, $\mathcal{U}_I$ and $\mathcal{U}_E$, problem \eqref{op:wet_g} is convex with respect to $\{R_i, \textbf{S}_i\}$ and can be solved using standard optimization techniques. The optimum solution can be found by solving problem \eqref{op:wet_g} for all possible combinations of user groups, that is, an exhaustive search should be implemented. Consider for example that $|\mathcal{U}_I| =4$ and $|\mathcal{U}_E|=4$ and that $K=10$. Then, problem \eqref{op:wet_g} (for a fixed $\mathcal{U}_I$ and $\mathcal{U}_E$) should be solved $\frac{K!}{|\mathcal{U}_I| !|\mathcal{U}_E|!(K-|\mathcal{U}_I| -|\mathcal{U}_E|)!} = 3.150$ times. Clearly, the optimum solution is impractical, even for a system with a small number of users. In that sense, any technique aside from the exhaustive search may be suboptimal. 

This fact motivates our second simplification: we decouple the decision of resource allocation and user grouping and propose a two-stage design strategy in which the user grouping is found based on suboptimal but less complex techniques. In other works, at the beginning of each frame, we first find the user groups $\mathcal{U}_I$ and $\mathcal{U}_E$, and then, for those fixed user groups, we solve the following convex optimization problem:
\begin{alignat}{3}
\mathop{\text{maximize}}_{\substack{\{R_i,\, \textbf{S}_i\}_{\forall i\in\mathcal{U}_I}}}& \quad \sum_{i\in\mathcal{U}_I} \omega_iR_i \label{op:wet1} \\
\textrm{subject to}
&  \quad C1\dots C6 \text{ of problem } \eqref{op:wet_g}.\nonumber
\end{alignat}
Note that due to $C5$, problem \eqref{op:wet1} is convex; otherwise, the objective function, i.e., the weighted sum rate, would not be convex due to the MUI.

In the next section, we are going to present a method to solve problem \eqref{op:wet1} for different settings. Later, in Section \ref{sec_sched}, we will present the user grouping techniques.

\section{Weighted Sum Rate Maximization with Harvesting Constraints}
\label{sr-ie}
The problem presented in \eqref{op:wet1} is convex and can be solved using numerical interior point methods \cite{boyd}. However, those methods usually have high computational complexity, and since we aim at finding a low-complexity solution, a customized algorithm should be developed. In some cases, it is possible to obtain the structure of the transmit covariance matrices in closed form and then develop an efficient algorithm based on that structure.  Unfortunately, it is not possible to find the closed-form expression of the optimal transmit covariances for the previous problem due to the constraint $C4$. However, as we will show later, it is possible to find the transmit covariance structure of problem \eqref{op:wet1} if $C4$ is not active.

To guarantee that constraint $C4$ is not active, we will assume that the set of information users is selected by the scheduler in a first stage in a way that they have enough battery such that $R^\star_i(t) < R_{\max,i}(C_i(t)), \, \forall i \in \mathcal{U}_I$ can be guaranteed in that particular scheduling period (later, we will comment on what to do in the unlikely event of violating the previous requirement). This is a reasonable assumption since users who have very low batteries should not be selected to receive information but to harvest energy. Due to the previous simplifying assumption, constraint $C4$ will not be active, and we therefore do not consider it in the optimization problem. This assumption considerably simplifies the resolution of the problem. 

Note that constraint $C5$ from the original problem \eqref{op:wet1} forces the precoder matrix $\textbf{B}_i$ to lie in the right null space of $\tilde{\textbf{H}}_i = [\textbf{H}_1^T \quad\dots \quad\textbf{H}_{i-1}^T \,\,\,\, \textbf{H}_{i+1}^T \quad\dots \quad\textbf{H}_N^T]^T\in\mathbb{C}^{(n_R-n_{R_i})\times n_T}$ \cite{spencer:04}. Computing the SVD of $\tilde{\textbf{H}}_i$ yields $\tilde{\textbf{H}}_i = \tilde{\textbf{U}}_i\mathbf{\tilde{\Lambda}}_i[\tilde{\textbf{V}}_i^{(1)} \quad \tilde{\textbf{V}}_i^{(0)}]^H$, where $\mathbf{\tilde{\Lambda}}_i$ is a diagonal matrix containing the singular values, and $\tilde{\textbf{V}}_i^{(0)}\in\mathbb{C}^{n_T\times(n_T-n_R+n_{R_i})}$ contains the right-singular vectors in the null space of $\tilde{\textbf{H}}_i$. Thus, $\textbf{B}_i$ can be written as $\textbf{B}_i = \tilde{\textbf{V}}_i^{(0)}\tilde{\textbf{B}}_i$ (with $\tilde{\textbf{B}}_i\in\mathbb{C}^{(n_T - n_R + n_{R_i})\times n_{S_i}})$, and then, $\textbf{S}_i = \tilde{\textbf{V}}_i^{(0)}\tilde{\textbf{S}}_i\tilde{\textbf{V}}_i^{(0)H}$, where $\tilde{\textbf{S}}_i = \tilde{\textbf{B}}_i\tilde{\textbf{B}}_i^H$. Now, the optimization problem can be rewritten in terms of the new optimization variables $\{\tilde{\textbf{S}}_i\}$. Let $\hat{\textbf{H}}_i = \textbf{H}_i\tilde{\textbf{V}}_i^{(0)}$ and $\hat{\textbf{H}}_{ji} = \textbf{H}_j\tilde{\textbf{V}}_i^{(0)}$. Note that if constraint $C4$ is not present in \eqref{op:wet1}, constraint $C3$ is tight at the optimum, i.e., $R^\star_i = \log \det \left(\textbf{I} + \hat{\textbf{H}}_i\tilde{\textbf{S}}^\star_i\hat{\textbf{H}}_i^H\right)$, and thus, the objective function is directly expressed as $\sum_{i\in\mathcal{U}_I} \omega_i\log \det \left(\textbf{I} + \hat{\textbf{H}}_i\tilde{\textbf{S}}_i\hat{\textbf{H}}_i^H\right)$. Then, problem \eqref{op:wet1} (without considering $C4$) is reformulated as
\begin{alignat}{3}
\mathop{\text{maximize}}_{\{\tilde{\textbf{S}}_i\}_{\forall i\in\mathcal{U}_I}}& \quad \sum_{i\in\mathcal{U}_I} \omega_i\log \det \left(\textbf{I} + \hat{\textbf{H}}_i\tilde{\textbf{S}}_i\hat{\textbf{H}}_i^H\right) \label{op:wet5} \\
\textrm{subject to}
&  \quad C1:\sum_{i\in\mathcal{U}_I}\Tr(\hat{\textbf{H}}_{ji}\tilde{\textbf{S}}_i\hat{\textbf{H}}_{ji}^H)  \ge Q_j, &\quad\quad \forall j \in \mathcal{U}_E \nonumber \\
&  \quad C2:\sum_{i\in\mathcal{U}_I}\Tr(\tilde{\textbf{S}}_i) + P_c^{tx}\le P_{\max}\nonumber\\
&  \quad C3:   \tilde{\textbf{S}}_i \succeq 0, & \forall i \in \mathcal{U}_I. \nonumber
\end{alignat}

The problem above can be checked to be convex since the objective function is concave and the constraints define a convex set. As a consequence, there exists a global optimal solution that can be obtained numerically by means of, for example, interior point methods \cite{boyd}. However, due to the fact that \eqref{op:wet5} is convex and satisfies Slater's conditions \cite{boyd}, the duality gap is zero, and the problem, therefore, can be solved using tools derived from the Lagrange duality theory, and the optimal structure of the transmit covariance matrices $\{\tilde{\textbf{S}}_i\}$ can be revealed. Let $\boldsymbol\lambda = \{\lambda_j\}_{j\in\mathcal{U}_E}$ be the vector of dual variables associated with constraint $C1$ and $\mu$ be the dual variable associated with constraint $C2$. The optimal solution of problem \eqref{op:wet5} is given by the following theorem in terms of $\boldsymbol\lambda^\star$ and $\mu^\star$.

\begin{theorem} \label{theorem1}
	The optimal solution of problem \eqref{op:wet5} has the following structure:
	\begin{equation}
	\tilde{\textbf{S}}_i^\star(\boldsymbol\lambda^\star,\mu^\star) = \textbf{A}_i^{-1/2}\hat{\textbf{V}}_i \hat{\textbf{D}}_i \hat{\textbf{V}}_i^H \textbf{A}_i^{-1/2},
	\label{optS}
	\end{equation}
	where matrix $\textbf{A}_i = \mu^\star \textbf{I} - \sum_{j\in\mathcal{U}_E} \lambda_j^\star \hat{\textbf{H}}_{ji}^H\hat{\textbf{H}}_{ji}$, $\hat{\textbf{V}}_i \in \mathbb{C}^{(n_T-n_R + n_{R_i})\times n_{S_i}}$ is obtained from the reduced SVD of matrix $\hat{\textbf{H}}_{i}^H\textbf{A}_i^{-1/2} = \hat{\textbf{U}}_i\hat{\mathbf{\Sigma}}_i^{1/2}\hat{\textbf{V}}_i^H$, with $\hat{\mathbf{\Sigma}}_i = \textrm{diag}(\hat{\sigma}_{1,i}, \dots , \hat{\sigma}_{n_{S_i},i})$, $\hat{\sigma}_{1,i} \ge \hat{\sigma}_{2,i} \ge \dots \ge \hat{\sigma}_{n_{S_i},i} > 0$, and $\hat{\textbf{D}}_i = \textrm{diag}(\hat{d}_{1,i}, \dots , \hat{d}_{n_{S_i},i})$, with $\hat{d}_{k,i} = (\omega_i/\log(2)-1/\hat{\sigma}_{k,i})^\infty_0$, $\forall i \in \mathcal{U}_I$ and $k = 1, \dots , n_{S_i}$.
\end{theorem}
\begin{proof}
	See Appendix \ref{app2}.
\end{proof}
Note the similarities in the precoder structure between the result presented in \eqref{optS} for the multiuser case and the result found in \cite{zhang:13b} for the single-user case. In the multiuser case, we have to find a set of multipliers associated with the per-user harvesting constraints, which makes the problem more complex to solve. Finally, the optimum data rate achieved by user $i$ is thus
\begin{equation}
R_i^\star = \log \det \left(\textbf{I} + \hat{\textbf{H}}_i\tilde{\textbf{S}}_i^\star\hat{\textbf{H}}_i^H\right) = \sum_{j=1}^{n_{S_i}} \log (1 + \hat{\sigma}_{j,i} \hat{d}_{j,i}), \quad \forall i\in\mathcal{U}_I.
\end{equation}

However, the above process is still pending the computation of the optimal dual variables since we assumed in the previous development that the dual variables were given (in Theorem \ref{theorem1}, matrix $\textbf{A}_i$ depends on the optimal values of the Lagrange multipliers). As long as we have a closed-formed expression of the covariance matrices $\tilde{\textbf{S}}_i(\boldsymbol\lambda,\mu)$ as a function of the dual variables, we can solve the dual problem of \eqref{op:wet5} by maximizing the dual function $g(\boldsymbol\lambda,\mu)$ subject to $\boldsymbol\lambda\succeq 0$, $\mu \ge 0$, and $\textbf{A}_i \succ 0\, \, \forall i$. This can be addressed by applying any subgradient-type method, such as, for example, the ellipsoid method \cite{bland:81}. It can be shown that the subgradient of $g(\boldsymbol\lambda, \mu)$, denoted as $\textbf{t}$, is given by $[\textbf{t}]_m = Q_{N+m} - \sum_{i\in\mathcal{U}_I}\Tr(\hat{\textbf{H}}_{(N+m)i}\tilde{\textbf{S}}_i\hat{\textbf{H}}_{(N+m)i}^H)$ for $1\le m\le M$ and $[\textbf{t}]_{M+1} = \Tr(\tilde{\textbf{S}}_i) - (P_{\max} - P^{tx}_c)$ \cite{bertsekas}, which represents the subgradient of $g(\boldsymbol\lambda, \mu)$ with respect to $\lambda_m$ and $\mu$, respectively ($[\textbf{t}]_{k}$ denotes the $k$-th entry of vector $\textbf{t}$), and $\tilde{\textbf{S}}_i$ is computed as in \eqref{optS} for a given $\boldsymbol\lambda$ and $\mu$ (for each step of the algorithm, we compute $\tilde{\textbf{S}}_i$ just by replacing, in expression \eqref{optS}, the optimal values of the Lagrange multipliers by their current values). Since the duality gap is zero, when we obtain the optimal dual variables ($\boldsymbol\lambda^\star$ and $\mu^\star$) with the ellipsoid method, the optimal solution $\tilde{\textbf{S}}_i^\star(\boldsymbol\lambda^\star,\mu^\star)$ converges to the primal optimal solution of problem \eqref{op:wet5}. As a summary, the algorithm that solves problem \eqref{op:wet5} is described in Table \ref{alg1} (this table was already presented in \cite{rub:13d} but is included here for the sake of completeness).

\begin{table}[t]
	\renewcommand{\arraystretch}{1.6}
	\caption{\textsc{Algorithm for Solving Problem \eqref{op:wet5}}}
	\label{alg1}
	\begin{center}
		\begin{tabular}{ l  l }
			\hline\hline
			1: & initialize $\boldsymbol\lambda\succeq 0$, $\mu \ge 0$ such that $\mu \textbf{I} - \sum_{j\in\mathcal{U}_E} \lambda_j \hat{\textbf{H}}_{ji}^H\hat{\textbf{H}}_{ji} \succ 0$, $\forall i$\\
			2: & \textbf{repeat}\\
			3: & \quad compute $\tilde{\textbf{S}}_i(\boldsymbol\lambda,\mu)$ $\forall i$ using \eqref{optS}\\
			4: & \quad compute subgradient of $g(\boldsymbol\lambda,\mu)$: \\
			5: & \quad \quad $[\textbf{t}]_m = Q_{N+m} - \sum_{i\in\mathcal{U}_I}\Tr(\hat{\textbf{H}}_{(N+m)i}\tilde{\textbf{S}}_i\hat{\textbf{H}}_{(N+m)i}^H)$ for $1 \le m \le M$ \\
			6: & \quad \quad $[\textbf{t}]_{M+1} = \Tr(\tilde{\textbf{S}}_i) - (P_{\max} - P^{tx}_c)$\\
			7: & \quad update $\boldsymbol\lambda$, $\mu$ using the ellipsoid method \cite{bland:81} subject to the following:\\
			& \quad \quad $\boldsymbol\lambda\succeq 0$, $\mu \ge 0$ and $\mu \textbf{I} - \sum_{j\in\mathcal{U}_E} \lambda_j \hat{\textbf{H}}_{ji}^H\hat{\textbf{H}}_{ji} \succ 0$, $\forall i$\\
			8: & \textbf{until} dual variables converge \\
			\hline\hline
		\end{tabular}
	\end{center}
\end{table}

\subsection{Particular Cases: Scenario with Only One Type of User}
\label{maxSR}
There exists a couple of particular cases of the problem presented before in which only one type of user is present in the system. Such simplified scenarios are found in real systems and will yield simpler optimization problems with lower computational complexity in the resolution of the resource allocation algorithm. For the sake of ease of readability of the paper, the mathematical developments of both particular cases have been moved to App. \ref{part_cases}.

\subsection{Tradeoff Analysis Between Weighted Sum Rate and Power Constraints}
\label{trade-off}
In this section, we analyze the multidimensional tradeoff between the objective function, that is, the weighted sum rate, and the set of power harvesting constraints. For simplicity, let us consider that $C_i(t)\,\,\forall i\in \mathcal{U}_I$ is high enough so that it could be assumed that $R_i^\star < R_{\max,i}$ and $R_i^\star = \log \det \left(\textbf{I} + \hat{\textbf{H}}_i\tilde{\textbf{S}}_i^\star\hat{\textbf{H}}_i^H\right)$. We would like to emphasize that, as the noise and channels are normalized, we will refer to the powers harvested by the receivers in terms of power units instead of Watts. Given this approach, we propose to use the \emph{Rate-Power} (R-P) region to characterize all the achievable sum rates (in bit/s/Hz) and power harvesting (in power units) $M+1$-tuples under a given power constraint as in \cite{zhang:13b}. The R-P region of problem \eqref{op:wet5} is defined as
\begin{eqnarray}
 & & \hspace{-4mm}\mathcal{C}_{\text{R-P}}((P_{\max} - P^{tx}_c), \{\omega_i\}) \triangleq \\ & & \Big\{\,(\textrm{SR};\{Q_j\}) \,\mid\, \exists\,\, \{\tilde{\textbf{S}}_i\} \,\, \text{with}\,\,\, \textrm{SR} \le \sum_{i\in\mathcal{U}_I}\omega_i \log \det \left(\textbf{I} + \hat{\textbf{H}}_i\tilde{\textbf{S}}_i\hat{\textbf{H}}_i^H\right), \nonumber \\
& &  \sum_{i\in\mathcal{U}_I}\Tr(\hat{\textbf{H}}_{ji}\tilde{\textbf{S}}_i\hat{\textbf{H}}_{ji}^H)\ge Q_j, \,\displaystyle\sum_{i\in\mathcal{U}_I}\Tr(\tilde{\textbf{S}}_i)+P^{tx}_c\le P_{\max}, \, \tilde{\textbf{S}}_i \succeq 0 \quad \forall j \in \mathcal{U}_E,\, \forall i \in \mathcal{U}_I \Big\}.\nonumber
\label{R-E}
\end{eqnarray}

\begin{figure}
	\centering
	\textbf{Three-dimensional R-P region}\\
	\includegraphics[width=0.6\textwidth]{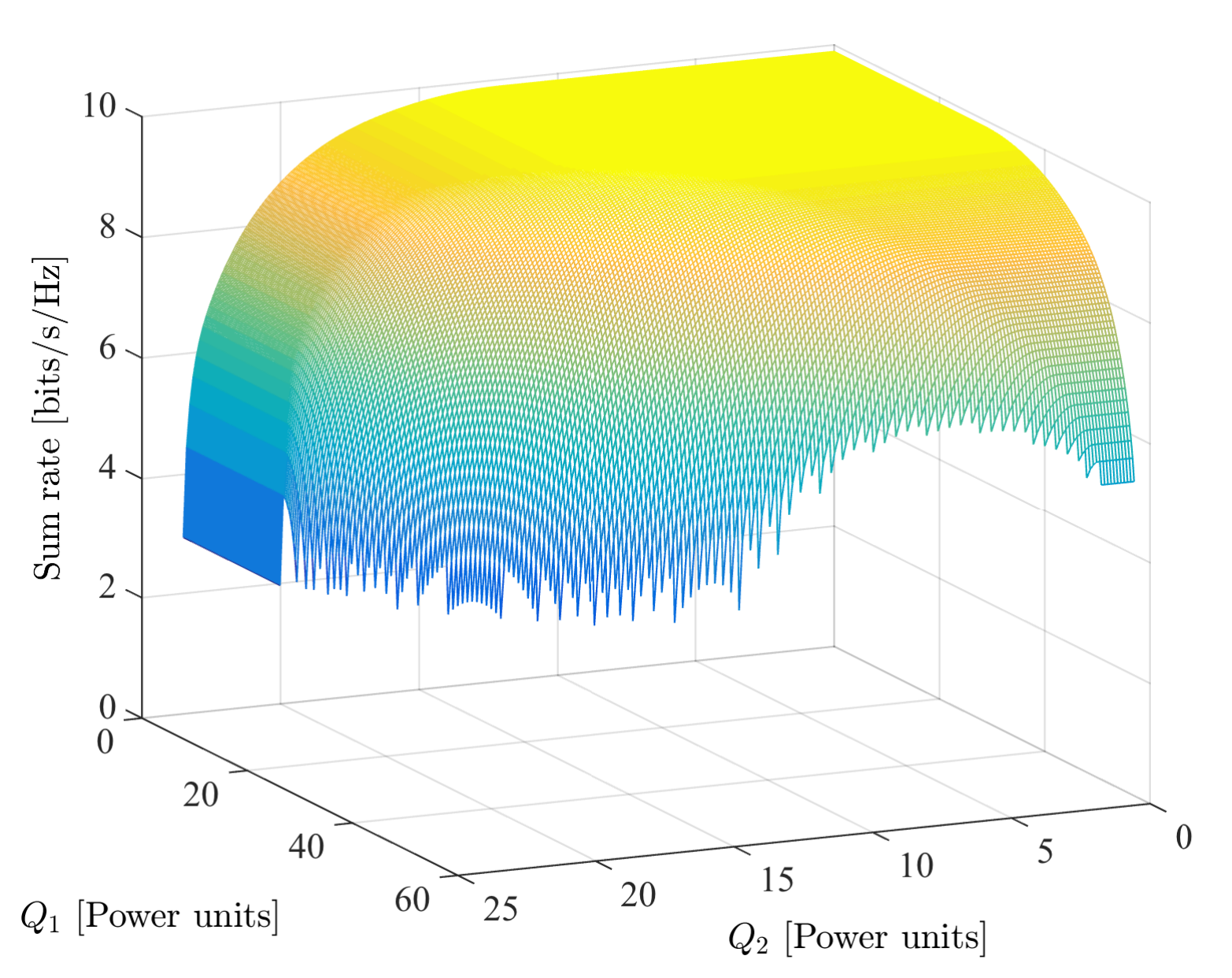}
	\caption{Representation of the three-dimensional R-P region of problem \eqref{op:wet5}. The figure represents the existing tradeoff between the optimal solution of the problem, i.e., the weighted sum rate, and the two power harvesting constraints.}
	\label{esq1}
\end{figure}

\begin{figure}
	\centering
	\textbf{Contour lines of the three-dimensional R-P region}\\
	\includegraphics[width=0.6\textwidth]{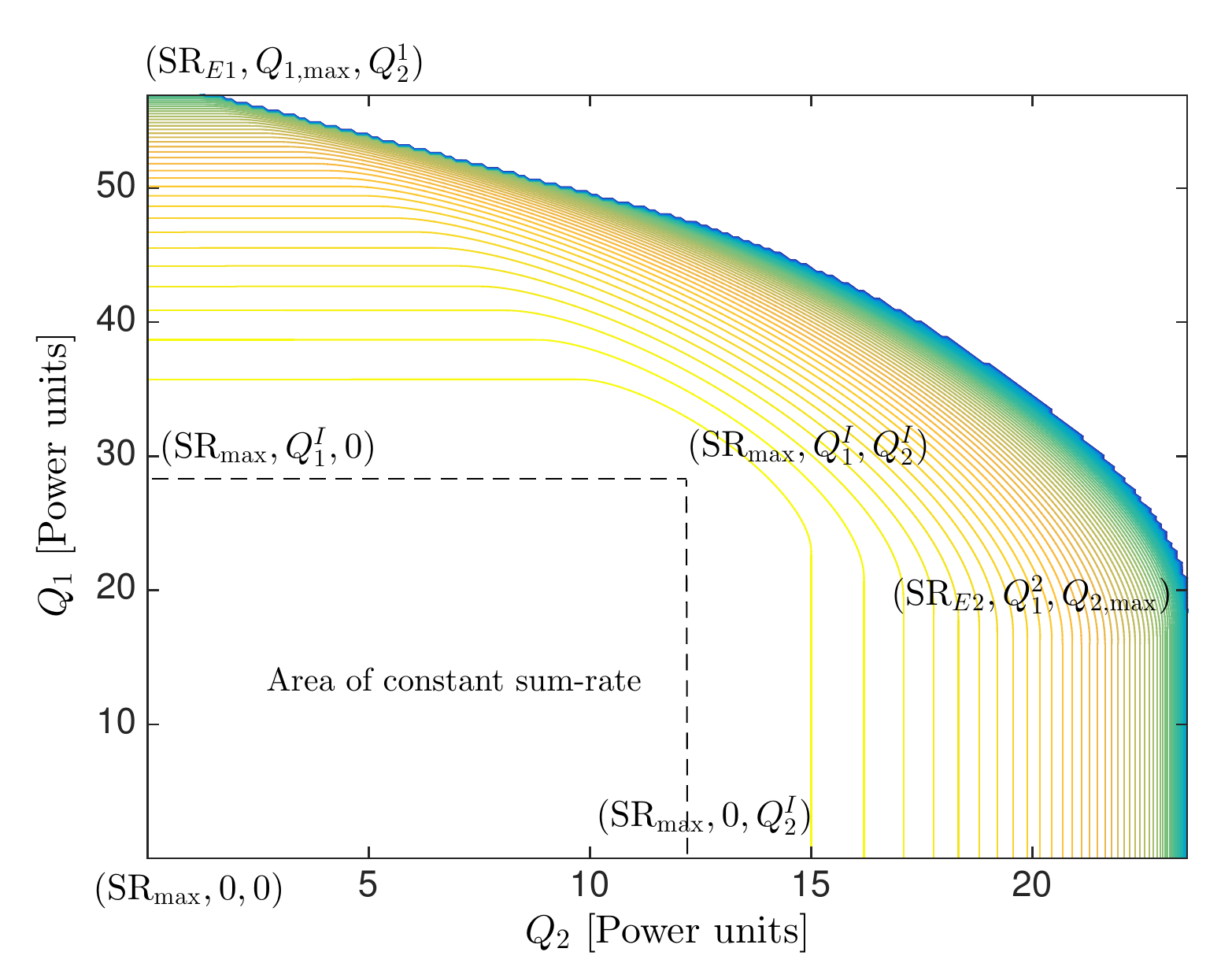}
	\caption{Contour lines of the three-dimensional R-P region of problem \eqref{op:wet5}. Note that the density of lines increases with the gradient of the surface, and the color indicates the value of such a surface. Note also that some important boundary characteristic points have been marked.}
	\label{esq2}
\end{figure}

To be able to graphically show an example of the tradeoff, we restrict the cardinality of the set of harvesting users and information users to be two, i.e., \mbox{$|\mathcal{U}_E| = 2$} and \mbox{$|\mathcal{U}_I| = 2$}, and for simplicity, we consider that $\omega_i = 1, \,\,\forall i\in\mathcal{U}_I$. In such a case, the tradeoff region between the sum rate and the two power constraints is a 3-dimensional surface. The setup taken as an example for this section is a BS with four transmit antennas and where all users have two antennas. The maximum transmission power at the BS is $P_{\max} - P^{tx}_c = 10$ W. The entries of the matrix channels are generated independently from a complex circularly symmetric Gaussian distribution with zero mean and variance equal to one.\footnote{The plots in Figs. \ref{esq1} and \ref{esq2} contain some of the results already shown in \cite{rub:13d}, which are included here for the sake of completeness.} 

Fig. \ref{esq1} depicts the 3-dimensional R-P region for the previous setup. As can be appreciated, the optimal sum rate solution is jointly concave on $Q_1$ and $Q_2$, as expected \cite{boyd}. The values of $Q_1$ and $Q_2$ for which the region is not defined correspond to situations where problem \eqref{op:wet5} is infeasible. To characterize the surface accurately, let us introduce the contour lines of the R-P region in Fig. \ref{esq2}. In the plot, when the lines are close together, the magnitude of the gradient is large. There are also some important boundary points marked in the 3-D plot of the surface. Those points can be computed in a simple way and provide us with useful cases that will be commented on in what follows.

Let us first start with the boundary point defined by $(\textrm{SR}_{\max},0,0)$. The power harvesting constraints for users 1 and 2 at this point are set to zero, and the solution of the problem therefore can be obtained from problem \eqref{op:wet2} (or from problem \eqref{op:wet5} with $Q_1 = Q_2 = 0)$. SR$_{\max}$ represents the maximum sum rate that can be achieved in this situation when no energy harvesting is imposed. The optimum covariance matrices were obtained in Section \ref{maxSR} and are denoted here as $\tilde{\textbf{S}}^\star_{\textrm{SR}_i}$ for the $i$-th user. Following that notation, the maximum sum rate can also be expressed as $\textrm{SR}_{\max} = \log \det \left(\textbf{I} + \hat{\textbf{H}}_1\tilde{\textbf{S}}^\star_{\textrm{SR}_1}\hat{\textbf{H}}_1^H\right) +\log \det \left(\textbf{I} + \hat{\textbf{H}}_2\tilde{\textbf{S}}^\star_{\textrm{SR}_2}\hat{\textbf{H}}_2^H\right)$.

Note that although when computing $\textrm{SR}_{\max}$, we do not apply power harvesting constraints, this does not necessarily mean that the actual harvested powers are zero. In this context, we have the boundary point $(\textrm{SR}_{\max},Q_1^\textrm{I},0)$, where $Q_1^\textrm{I}$ represents the power harvested by user 1 when the precoder matrices are the ones that maximize the weighted sum rate, i.e., $Q_1^\textrm{I} = \Tr(\hat{\textbf{H}}_{11}\tilde{\textbf{S}}^\star_{\textrm{SR}_1}\hat{\textbf{H}}_{11}^H) + \Tr(\hat{\textbf{H}}_{12}\tilde{\textbf{S}}^\star_{\textrm{SR}_2}\hat{\textbf{H}}_{12}^H)$. The same can be said for the boundary point $(\textrm{SR}_{\max},0,Q_2^\textrm{I})$, where $Q_2^\textrm{I} = \Tr(\hat{\textbf{H}}_{21}\tilde{\textbf{S}}^\star_{\textrm{SR}_1}\hat{\textbf{H}}_{21}^H) + \Tr(\hat{\textbf{H}}_{22}\tilde{\textbf{S}}^\star_{\textrm{SR}_2}\hat{\textbf{H}}_{22}^H)$. Then, there is a fourth point that defines a flat surface (or tableland) of constant sum rate $\textrm{SR}_{\max}$, which is the combination of the two previous points, $(\textrm{SR}_{\max},Q_1^\textrm{I},Q_2^\textrm{I})$. In other words, the tableland of constant maximum weighted sum rate $\textrm{SR}_{\max}$ defines all possible values of harvested power constraints for which constraints $C1$ are not active and thus do not affect the optimum value of the weighted sum rate.

Now, let us consider the boundary points in terms of maximum harvested power. On top of the figure, there is the point $(\textrm{SR}_{\textrm{E}1}, Q_{1,\max}, Q_2^1)$. This point corresponds to the situation in which the power harvested by user 1 is a maximum or, in other words, the maximum value of $Q_1$ for which problem \eqref{op:wet5} is feasible, assuming no constraint on the power to be harvested by user 2. To calculate $Q_{1,\max}$, we solve the following optimization problem:
\begin{alignat}{2}
\mathop{\text{maximize}}_{\tilde{\textbf{S}}_{\textrm{E}1}}& \quad \Tr(\hat{\textbf{H}}_{11}\tilde{\textbf{S}}_{\textrm{E}1}\hat{\textbf{H}}_{11}^H) \label{op:wet6} \\
\textrm{subject to}
&  \quad C1:\Tr(\tilde{\textbf{S}}_{\textrm{E}1}) + P^{tx}_c\le P_{\max}\nonumber\\
&  \quad C2:  \tilde{\textbf{S}}_ {\textrm{E}1}\succeq 0, \nonumber
\end{alignat}
where $\tilde{\textbf{S}}_{\textrm{E}1}$ represents the sum of the two covariance matrices for the information users (note that in this problem, the objective function and the constraint depend on such matrices through their sum), and the objective function is the power harvested by user 1. Now, by applying the result from Proposition \ref{prop_eig}, we obtain the solution of problem \eqref{op:wet6} as follows. Let the reduced eigen-decomposition of $\hat{\textbf{H}}_{11}^H\hat{\textbf{H}}_{11}$ be $\hat{\textbf{U}}_{11}\hat{\mathbf{\Lambda}}_{11}\hat{\textbf{U}}_{11}^H$ such that $\hat{\textbf{u}}_{11,\max}$ is the eigenvector associated with the maximum eigenvalue $\hat{\lambda}_{11,\max}$. Then, the solution to the previous problem is based on the following inequality: $\Tr(\tilde{\textbf{S}}_{\textrm{E}1}\hat{\textbf{H}}_{11}^H\hat{\textbf{H}}_{11}) \le \hat{\lambda}_{11,\max}\Tr(\tilde{\textbf{S}}_{\textrm{E}1}) = \hat{\lambda}_{11,\max} \times(P_{\max} - P^{tx}_c)$ (since at the optimum $\Tr(\tilde{\textbf{S}}^\star_{\textrm{E}1}) = P_{\max} - P^{tx}_c)$, where such inequality becomes equality if $\tilde{\textbf{S}}^\star_{\textrm{E}1} = (P_{\max} - P^{tx}_c)\times \hat{\textbf{u}}_{11,\max}\hat{\textbf{u}}_{11,\max}^H$. In this case, the maximum harvested energy is accomplished by \emph{energy beamforming}\footnote{The concept of energy beamforming was already introduced in \cite{zhang:13b}.} (i.e., rank 1) to the best eigenmode of the equivalent channel $\hat{\textbf{H}}_{11}^H\hat{\textbf{H}}_{11}$. Then, we obtain $Q_{1,\max} = \Tr(\hat{\textbf{H}}_{11}\tilde{\textbf{S}}^\star_{\textrm{E}1}\hat{\textbf{H}}_{11}^H) = (P_{\max} - P^{tx}_c)\times\hat{\lambda}_{11,\max}$. According to this, the weighted sum rate obtained by solving problem \eqref{op:wet5} and $Q_1 = Q_{1,\max}, Q_2 = 0$ (denoted as $\textrm{SR}_{\textrm{E}1}$) is $\textrm{SR}_{\textrm{E}1} = \log \det \left(\textbf{I} + \hat{\textbf{H}}_1\tilde{\textbf{S}}^\star_{\textrm{E}1}\hat{\textbf{H}}_1^H\right) + \log \det \left(\textbf{I} + \hat{\textbf{H}}_2\tilde{\textbf{S}}^\star_{\textrm{E}1}\hat{\textbf{H}}_2^H\right)$. Note that, even though we do not apply the power harvesting constraint of user 2 when computing $\tilde{\textbf{S}}_{\textrm{E}1}$, it does not mean that the actual power harvested by user 2 is zero. In this context, we define the last coordinate of the point, denoted as $Q_2^1$, which represents the power harvested by user 2 when the covariance matrix is $\tilde{\textbf{S}}^\star_{\textrm{E}2}$, i.e., $Q_2^1 = \Tr(\hat{\textbf{H}}_{21}\tilde{\textbf{S}}^\star_{\textrm{E}2}\hat{\textbf{H}}_{21}^H)$. The same reasoning can be applied to obtain the last boundary point $(\textrm{SR}_{\textrm{E}2}, Q_1^2, Q_{2,\max})$ by interchanging the roles of users 1 and 2.

The remaining boundary points in the curve can be obtained by properly varying the values of $Q_1$ and $Q_2$ ($0 \le Q_1\le Q_{1,\max}$, $0 \le Q_2 \le Q_{2,\max}$) in problem \eqref{op:wet5}.

\section{User Selection Policies}
\label{sec_sched}
Thus far, we have assumed that the two groups of users, i.e., $\mathcal{U}_I$ and $\mathcal{U}_E$, were known. The goal of this section is to propose a grouping strategy to select which users should go into each set in a way that the aggregated throughput over time is maximized. As the channels and batteries fluctuate throughout time, the users in each group may also change from frame to frame. In this section, we will assume that the values of $\{Q_j\}$ are known and fixed. The management of these values is beyond the scope of the paper (see the work in \cite{rub:15a}, where the authors propose some procedures to adjust the values of $\{Q_j\}$, considering the impact on the system performance).   

As previously noted, the optimal information and harvesting grouping should be obtained by joint exhaustive search (see Section \ref{sec_joint}). This search is prohibitively complex, and suboptimum techniques therefore should be derived. The case of having only information users has been studied in the literature, and suboptimal techniques that perform close to the optimum one have been proposed \cite{dimic:05}, \cite{sigdel:09}. In this paper, to keep the overall complexity as low as possible without compromising the performance of the system, we present suboptimal techniques for the user grouping for both kinds of users, i.e., information and harvesting users. This is one of the major contributions of our paper, that is, work with users that have different objectives. Additionally, as we will show, the proposed greedy algorithms take into account that the selection of the harvesting users impacts directly the performance of the information users, that is, there is a coupling behavior between both aspects.

The overall user grouping strategy will be divided into two stages. In the first stage (that will be known as \emph{super-grouping}), we will provide a preselection of user candidates to be in each set. This will depend primarily on the current energies available at the batteries, and it will be run at a longer time scale, every few scheduling periods or frames. For the second stage, known as \emph{grouping}, we are going to present two different user grouping strategies that will be run at every frame. The strategy with the highest complexity provides a better performance than the simpler strategy. 

In the first (simpler) approach, we will split the user grouping further into two stages. The first stage selects the information users, $\mathcal{U}_I$, from the super-grouping set $\mathcal{U}^S_I$ based on a greedy approach, whereas the second stage selects the harvesting users, $\mathcal{U}_E$, based on the already selected information users. In the second approach, we will develop a joint information-harvesting grouping strategy, which constitutes an intermediate approach between the first simple approach and the optimum approach based on exhaustive search.

\subsection{User Supergrouping Strategy}
Recall that when we derived the optimal precoder matrix in Section \ref{sr-ie}, we assumed that the optimal rates would fulfill $R^\star_i(t) < R_{\max,i}(t), \, \forall i \in \mathcal{U}_I$ for any particular frame, and therefore, constraints $C4$ in problem \eqref{op:wet1} were not active. This is achieved by preselecting the users that are to be scheduled for data transmission or battery charging. In our proposed approach, we first implement a selection of candidates to be in $\mathcal{U}_I$ and $\mathcal{U}_E$, known as $\mathcal{U}^S_I$ and $\mathcal{U}^{S}_E$, such that $\mathcal{U}_I\subseteq \mathcal{U}^S_I$, $\mathcal{U}_E\subseteq \mathcal{U}^S_E$, and $|\mathcal{U}^S_I| + |\mathcal{U}^S_E| = K$, and we then select the users that finally go into the sets $\mathcal{U}_I$ and $\mathcal{U}_E$. The proposed supergrouping algorithm is presented in Table \ref{algSF} and works as follows: we set a threshold $\alpha$ such that $0\le\alpha\le 1$. Then, we compute the ratio of the current battery level and the battery capacity for all users, and we then order these ratios increasingly. If the middle ratio of the previous list is greater than the value of the threshold $\alpha$, we then split the overall group by half and put half of the users in $\mathcal{U}^S_I$ and the other half in $\mathcal{U}^S_E$. On the other hand, if the middle ratio of the previous list is lower than the value of $\alpha$, we find the user with battery ratio closest to the value of $\alpha$ and put all users with lower ratios than the one closest to $\alpha$ in the harvesting set and the remaining users in the information set. The larger the value of $\alpha$, the greater the number of users that will be included in the harvesting set $\mathcal{U}^S_E$. Note that the BS has to know the battery levels of all users, which implies that receivers must send the battery levels through a feedback channel and, hence, the battery levels must be quantized (in \cite{rub:14a}, we addressed the problem of quantizing the battery levels and evaluated the effect on the overall system performance, and we conclude that a few bits for quantization is enough to obtain good performance). 

\begin{table}[!t]
	\renewcommand{\arraystretch}{1.2}
	\caption{\textsc{Algorithm to obtain the super-frame sets $\mathcal{U}^S_I$ and $\mathcal{U}^S_E$}}
	\label{algSF}
	\begin{center}
		\begin{tabular}{ r  l }
			\hline
			\hline
			1:& set a threshold $0\le \alpha \le 1$\\
			2:& order the users increasingly with the following rule:\\
			& $\frac{C_1(t)}{C^1_{\max}} \le \frac{C_2(t)}{C^2_{\max}} \le \dots \le \frac{C_{K/2}(t)}{C^{K/2}_{\max}} \le \frac{C_{K/2+1}(t)}{C^{K/2+1}_{\max}}\le\dots \le \frac{C_{K}(t)}{C^{K}_{\max}}$\\
			3:& \textbf{if} $\alpha <  \frac{C_{K/2}(t)}{C^{K/2}_{\max}}$\\
			4:& \quad \quad users $\{1, 2, \dots, K/2\}$ go to $\mathcal{U}^S_E$\\
			5:& \quad \quad users $\{K/2+1, K/2+2, \dots, K\}$ go to $\mathcal{U}^S_I$\\
			6:&\textbf{else}\\
			7:& \quad \quad find the user $m$ such that $m=\arg\min_{i} \left|\frac{C_i(t)}{C^i_{\max}} - \alpha\right|$\\
			8:& \quad \quad users $\{1, 2,\dots, m\}$ go to $\mathcal{U}^S_E$\\
			9:& \quad \quad users $\{m+1, m+2, \dots, K\}$ go to $\mathcal{U}^S_I$ \\
			10:& \textbf{end if}\\
			\hline
			\hline
		\end{tabular}
	\end{center}
\end{table}

\subsection{Disjoint Information and Harvesting User Grouping}
\label{deco_hs}
This first approach is based on two stages. In the first stage, the selection of the information users follows a greedy approach, in which each user is added at a time and the maximization of the weighted sum rate without harvesting constraints is evaluated for all possible candidate information users with the already selected users. No harvesting users are considered at this stage. 

Let us assume, for simplicity, that every information user has the same number of antennas, i.e., $n_{R_i} = N_R, \,\forall i\in \mathcal{U}_T$. The maximum number of simultaneous users to be served following the BD strategy is then $U = \lceil \frac{n_T}{N_R}\rceil$ \cite{spencer:04}. The algorithm for selecting the information users is shown in Table \ref{alg2}; first, we select the user that can achieve the greatest weighted rate\footnote{A way to calculate the weights $\omega_i$ can be based on the achieved average rate as in the proportional fair (PF) scheme \cite{jalali:00}, \cite{wang:07}, \cite{liu:10}. In that case, the weights are computed as $\omega_i (t)= \frac{1}{T_i(t)}$, being that $T_i(t)$ is the exponentially averaged rate calculated as $T_i(t) = \left(1-\frac{1}{T_c}\right)T_i(t-2) + \frac{1}{T_c} R_i(t-1)$, where $T_c$ is the effective length of the impulse response of the exponential averaging filter, and $R_i(t-1)$ is the rated assigned to the $i$-th user in the $(t-1)$-th frame. Note that if the $i$-th user was not selected to be in $\mathcal{U}_I$ during the $(t-1)$-th frame, then $R^\star_i(t-1) = 0$. Otherwise, $R_i(t-1) =R^\star_i(t-1)$, i.e., the rate $R_i(t-1)$ corresponds to the solution of problem \eqref{op:wet5} during the $(t-1)$-th frame. Note that many other fairness criteria could be introduced by properly adjusting the weights.}. Then, we incorporate one user at a time into the set only if the accumulated weighted sum rate increases due to incorporating such a user (weighted sum rate evaluated with the already selected users). The algorithm ends when there is no improvement in the weighted sum rate or when the maximum number of users to be scheduled ($U$) is reached.

\begin{table}[!t]
	\renewcommand{\arraystretch}{1.2}
	\caption{\textsc{Algorithm to obtain the set of information users $\mathcal{U}_I$}}
	\label{alg2}
	\begin{center}
		\begin{tabular}{ r  l }
			\hline
			\hline
			1:& set $\mathcal{U}_I = \emptyset$, $Q_i\ge 0$, and $\omega_i>0, \, \forall i \in\mathcal{U}_T$ \\
			2:& find $i_1 = \arg\max_{\forall i \in \mathcal{U}^S_I} \max_{\textbf{S}_i}\omega_i\log\det\left(\textbf{I} + \textbf{H}_i\textbf{S}_i\textbf{H}_i^H\right)$\\
			& subject to $\Tr(\textbf{S}_i)\le P_T$, $\textbf{S}_i\succeq 0$\\ 
			3:& set $f_{\textrm{temp}} = \omega_{i_1}\log\det(\textbf{I} + \textbf{H}_{i_1}\textbf{S}_{i_1}\textbf{H}_{i_1}^H)$\\
			4:& set $\mathcal{U}_I \leftarrow \mathcal{U}_I \cup \{i_1\}, \,\, \mathcal{U}^S_I \leftarrow \mathcal{U}^S_I \setminus \{i_1\}$\\
			5:& \textbf{for} $j = 2$ to $U$\\
			6:& \quad \, \textbf{for} every $i \in \mathcal{U}^S_I$\\
			7:& \quad \quad\,   let $\mathcal{U}^{(i)}_I = \mathcal{U}_I \cup \{i\}$\\
			8:& \quad \quad\, solve \eqref{op:wet5} without $C1$, and obtain $R^\star_m, \, \forall m\in\mathcal{U}^{(i)}_I$\\
			9:&\quad \quad\,   compute $f_i = \sum_{m\in\mathcal{U}^{(i)}_I} \omega_mR^\star_m$\\
			10:& \quad \, \textbf{end for}\\
			11:& \quad \, let $i_j = \arg \max_{i \in \mathcal{U}^S_I} f_i$\\
			12:& \quad \, \textbf{if} $f_{i_j} < f_{\textrm{temp}} \longrightarrow$, go to 17 (break for)\\
			13:& \quad \, \textbf{else} \\
			14:& \quad \quad \, $\mathcal{U_I} \leftarrow \mathcal{U}_I \cup \{i_j\}, \,\, \mathcal{U}^S_I \leftarrow \mathcal{U}^S_I \setminus \{i_j\}$\\
			15:& \quad \quad \, let $f_{\textrm{temp}} = f_{i_j}$\\
			16:& \quad \, \textbf{end if}\\
			17:& \textbf{end for}\\
			\hline
			\hline
		\end{tabular}
	\end{center}
\end{table}

Note that the distances from the BS to the users are taken into account implicitly in the algorithm since, in step 2 and step 8 of Table \ref{alg2}, we select users according to the rates. These rates depend on the channel matrices $\{\textbf{H}_i\}$, and the components of these matrices, of course, will be small if the distances are large. Therefore, the distances will have a direct impact on the selection of users.

Once we have selected the information users, we continue with the selection of the harvesting users in the second stage of this grouping strategy. The idea is to select the harvesting users so that when the resource allocation strategy is executed, they affect (reduce) the system performance as little as possible (see Section \ref{trade-off}). Let $\textbf{S}^\star = \sum_{i\in\mathcal{U}_I}\textbf{S}^\star_i$, where $\mathcal{U}_I$ and $\{\textbf{S}^\star_i\}_{i\in\mathcal{U}_I}$ are the information user set and the optimum covariance matrices obtained from the algorithm detailed in Table \ref{alg2}, respectively. The algorithm works as follows. For each harvesting user $j$, we evaluate and decreasingly order $\Tr(\textbf{H}_j\textbf{S}^\star\textbf{H}_j) - Q_j$ and select the first $M$ harvesting users according to this order. Note that in the previous expression, we are evaluating how the optimum covariance matrices of the selected information users transmit power in the geometrical direction of the channels of the harvesting users. We also take into account the minimum required power to be harvested $Q_j$ to ensure feasibility of the solution of the resource allocation problem. The algorithm is presented in Table \ref{alg3}.

\begin{table}[!t]
	\renewcommand{\arraystretch}{1.2}
	\caption{\textsc{Algorithm to obtain the set of harvesting users $\mathcal{U}_E$}}
	\label{alg3}
	\begin{center}
		\begin{tabular}{ r  l }
			\hline
			\hline
			1:& input: $\mathcal{U}_I$ taken from algorithm in Table \ref{alg2}, $\textbf{S}^\star= \sum_{i\in\mathcal{U}_I}\textbf{S}^\star_i,$ \\
			2:& evaluate $m_j  = \Tr(\textbf{H}_j\textbf{S}^\star\textbf{H}_j) - Q_j,\,\,\forall j \in \mathcal{U}^S_E$\\
			3:& decreasingly order $m_j$\\
			4:& construct $\mathcal{U}_E$ with the users corresponding to the first $M$ ordered terms of $m_j$\\ 
			\hline
			\hline
		\end{tabular}
	\end{center}
\end{table}

\subsection{Joint Information and Harvesting User Grouping}
\label{cou_hs}
In this second approach, the selection of the information and harvesting users is coupled. Due to this joint approach, the system performance will be degraded less by the effect of having harvesting users in the system compared with the previous decoupled approach. However, the computational complexity increases as more combinations need to be evaluated. 

The algorithm for selecting the information users is based on the same greedy approach that we presented before. The difference is that, now, instead of selecting the information users and then the harvesting users, we select both types of users simultaneously. For simplicity in the formulation, let us consider that $M$ is an integer multiple of $U$ and define $k = \frac{M}{U}$ (we will comment later on how we could apply the algorithm if that was not the case). The idea behind the algorithm is as follows. We select one information user $q$ and obtain its optimum covariance matrix $\textbf{S}^\star_q$. Then, we find the best $k$ harvesting users based on the principle developed in Table \ref{alg3}. After that, we select another information user and repeat the same process until there is no improvement in the objective function. Due to the fact that the grouping is coupled, we consider the impact of having selected harvesting users on the future selection of information users. The specific details of the joint algorithm are presented in Table \ref{alg4}.
\begin{table}[t]
	\renewcommand{\arraystretch}{1.2}
	\caption{\textsc{Algorithm to jointly obtain the set of information and harvesting users $\mathcal{U}_I$, $\mathcal{U}_E$}}
	\label{alg4}
	\begin{center}
		\begin{tabular}{ r  l }
			\hline
			\hline
			1:& set $\mathcal{U}_I = \emptyset$, $Q_i\ge 0$, and $\omega_i>0, \, \forall i \in\mathcal{U}_T$ \\
			2:& find $i_1 = \arg\max_{\forall i \in \mathcal{U}^S_I} \max_{\textbf{S}_i}\omega_i\log\det\left(\textbf{I} + \textbf{H}_i\textbf{S}_i\textbf{H}_i^H\right)$\\
			& subject to $\Tr(\textbf{S}_i)\le P_T$, $\textbf{S}_i\succeq 0$\\ 
			3:& set $f_{\textrm{temp}} = \omega_{i_1}\log\det(\textbf{I} + \textbf{H}_{i_1}\textbf{S}_{i_1}\textbf{H}_{i_1}^H)$\\
			4:& set $\mathcal{U}_I \leftarrow \mathcal{U}_I \cup \{i_1\}, \,\, \mathcal{U}^S_I \leftarrow \mathcal{U}^S_I \setminus \{i_1\}$\\
			5:& evaluate $m_j  = \Tr(\textbf{H}_j\textbf{S}^\star_{i_1}\textbf{H}_j) - Q_j,\,\,\forall j \in \mathcal{U}^S_E$\\
			6:& find the $k$ users with highest value of $m_j$. Put them in set $\mathcal{H}$\\
			7:& set $\mathcal{U}_E \leftarrow \mathcal{U}_E \cup \mathcal{H}, \,\, \mathcal{U}^S_E \leftarrow \mathcal{U}^S_E \setminus \mathcal{H}$, $\mathcal{H} = \emptyset$\\
			8:& \textbf{for} $j = 2$ to $U$\\
			9:& \quad \, \textbf{for} every $i \in \mathcal{U}^S_I$\\
			10:& \quad \quad\,   let $\mathcal{U}^{(i)}_I = \mathcal{U}_I \cup \{i\}$\\
			11:& \quad \quad\, solve \eqref{op:wet5}, and obtain $R^\star_m, \,\textbf{S}^\star_m,\, \forall m\in\mathcal{U}^{(i)}_I$\\
			12:&\quad \quad\,   compute $f_i = \sum_{m\in\mathcal{U}^{(i)}_I} \omega_mR^\star_m$\\
			13:& \quad \, \textbf{end for}\\
			14:& \quad \, let $i_j = \arg \max_{i \in \mathcal{U}^S_I} f_i$\\
			15:& \quad \, \textbf{if} $f_{i_j} < f_{\textrm{temp}} \longrightarrow$, go to 23 (break for)\\
			16:& \quad \, \textbf{else} \\
			17:& \quad \quad \, $\mathcal{U_I} \leftarrow \mathcal{U}_I \cup \{i_j\}, \,\, \mathcal{U}^S_I \leftarrow \mathcal{U}^S_I \setminus \{i_j\}$\\
			18:& \quad \quad \, let $f_{\textrm{temp}} = f_{i_j}$\\
			19:& \quad \, \textbf{end if}\\
			20:& \quad \, evaluate $m_j  = \Tr(\textbf{H}_j\sum_{i\in\mathcal{U}_I}\textbf{S}^\star_i\textbf{H}_j) - Q_j,\,\,\forall j \in \mathcal{U}^S_E$\\
			21:& \quad \, find the $k$ users with highest value of $m_j$. Put them in set $\mathcal{H}$\\
			22:& \quad \, set $\mathcal{U}_E \leftarrow \mathcal{U}_E \cup \mathcal{H}, \,\, \mathcal{U}^S_E \leftarrow \mathcal{U}^S_E \setminus \mathcal{H}$, $\mathcal{H} = \emptyset$\\
			23:& \textbf{end for}\\
			\hline
			\hline
		\end{tabular}
	\end{center}
\end{table}

The main difference with the algorithm in Table \ref{alg3} is that, now, we solve problem \eqref{op:wet5} with constraints $C1$, that is, with harvesting users, which increases the complexity of the overall grouping procedure. As addressed before, if $M$ is not an integer multiple of $U$, we can introduce more harvesting users in step 21 in Table \ref{alg4} in some iterations, e.g., if $M = 7$ and $U=3$, we first select 3 harvesting users and then 2 harvesting users in the other 2 iterations.

\section{Overall User Grouping and Resource Allocation Algorithm}\label{sec_overall_algorithm}

In the following, we present a summary of the overall algorithm that consists of the user supergrouping, the user grouping, and the resource allocation stages presented in the previous two sections. Note that the user supergrouping is carried out every few frames, whereas the user grouping is executed at each frame. If, for some reason, the supergrouping algorithm fails in fulfilling $R^\star_i(t) < R_{\max,i}(t), \, \forall i \in \mathcal{U}_I$ (an event that would be unlikely to happen), then for those users for which $R^\star_i(t) \ge R_{\max,i}(t)$, we just transmit information in some channel accesses of the frame until their battery is over. The overall algorithm is detailed in Table \ref{algF}.

\begin{table}[!t]
	\renewcommand{\arraystretch}{1.2}
	\caption{\textsc{Overall user grouping and resource allocation algorithm}}
	\label{algF}
	\begin{center}
		\begin{tabular}{ r  l }
			\hline
			\hline
			& \underline{beginning of a super-frame:}\\
			1:& \quad run \textbf{user supergrouping} algorithm in Table \ref{algSF}: obtain sets $\mathcal{U}^S_I$ and $\mathcal{U}^S_E$\\
			& \underline{beginning of each frame (two options):}\\
			& \emph{option 1:}\\
			2a:&  \quad run \textbf{information user grouping} algorithm in Table \ref{alg2}: obtain set $\mathcal{U}_I$\\
			2b:& \quad run \textbf{harvesting user grouping} algorithms in Table \ref{alg3}: obtain set $\mathcal{U}_E$\\
			2c:&\quad run \textbf{resource allocation} algorithm in Table \ref{alg1}\\
			&\emph{option 2:}\\
			3a:&\quad  run \textbf{joint information and harvesting grouping} algorithm in Table \ref{alg4}:\\
			&\quad obtain sets $\mathcal{U}_I$ and $\mathcal{U}_E$\\
			3b:& \quad run \textbf{resource allocation} algorithm in Table \ref{alg1}\\
			& \underline{end of each frame:}\\
			4:& update batteries:\\
			& \quad \quad $C_i(t) =  \left(C_i(t-1) - T_f P_{\textrm{tot},i}^{r_x}(R^\star_i(t-1))\right)_0^{C_{\max}^i}$, $\quad \forall i \in \mathcal{U}_I$\\
			& \quad \quad $C_j(t) = \left(C_j(t-1) + T_f\bar{Q}_j(t-1) - T_f P_{c}^{r_x}\right)_0^{C_{\max}^j}$, $\quad \forall j \in \mathcal{U}_E$\\
			5:& update weights (e.g., using a PF approach):\\
			& \quad \quad $w_i(t) = \frac{1}{T_i(t)}$, \quad $T_i(t) = \left(1-\frac{1}{T_c}\right)T_i(t-2) + \frac{1}{T_c} R^\star_i(t-1)$\\
			\hline
			\hline
		\end{tabular}
	\end{center}
\end{table}

\section{Results and Discussion}
\label{num_sec}
In this section, we perform some numerical analysis of the proposed grouping and resource allocation strategies. The system comprises one transmitter with 8 antennas and 30 users $(|\mathcal{U}_T| = 30)$ with 2 antennas each. The maximum radiated power is $P_{\max} = 11$ W, and the transmitter front-end consumption is $P^{tx}_c = 1$ W. Front-end power consumption at the receiver is $P^{rx}_c = 100$ mW, and the model used for decoding is exponential, i.e., $P_{\text{dec}}(R) = c_1 \text{e}^{c_2R}$, where $c_1 = 30$, and $c_2 = 0.75$ \cite{rub:14a}. The frame duration is equal to $T_f = 100$ ms, and the super-frame duration is equal to 3 s. The channel matrices are generated randomly with i.i.d. entries distributed according to $\mathcal{CN} (0,1)$. The noise power is normalized to 1. The effective window length for the PF scheme is $T_c = 5$. The percentage used for supergrouping is $\alpha = 0.1$. The battery capacities are generated randomly from 3,000 to 10,000 energy units. As we mentioned previously, we assume that all the harvesting constraints are the same for all users and fixed for all periods to $Q_j = 50$ power units, unless stated otherwise. A strategy on how to manage and dynamically adjust the values of the $\{Q_j\}$ was proposed in \cite{rub:15a} and is beyond the scope of this paper. 

In the simulations, we compare our proposed two methods with two other schemes. As there are no proposals in the literature for user scheduling in the SWIPT framework, we compare our approaches with traditional schemes. In one of the schemes, we assume that the supergrouping and grouping are implemented with a round robin strategy. We will denote this strategy RR-SF/RR-F. In the other scheme, we consider that random selection of users is implemented at both levels as well. This strategy will be denoted by Ra-SF/Ra-F. On the other hand, the proposed supergrouping strategy (Table \ref{algSF}) will be denoted by LB, and the grouping will be denoted according to the algorithm: DHS for the decoupled approach presented in Section \ref{deco_hs} (Tables \ref{alg2} and \ref{alg3}) and CHS for the approach presented in Section \ref{cou_hs} (Table \ref{alg4}). 

\subsection{Time Evolution Simulations}
\begin{figure}
	\centering
	\textbf{Time evolution of the battery levels}\\
	\includegraphics[width=0.6\textwidth]{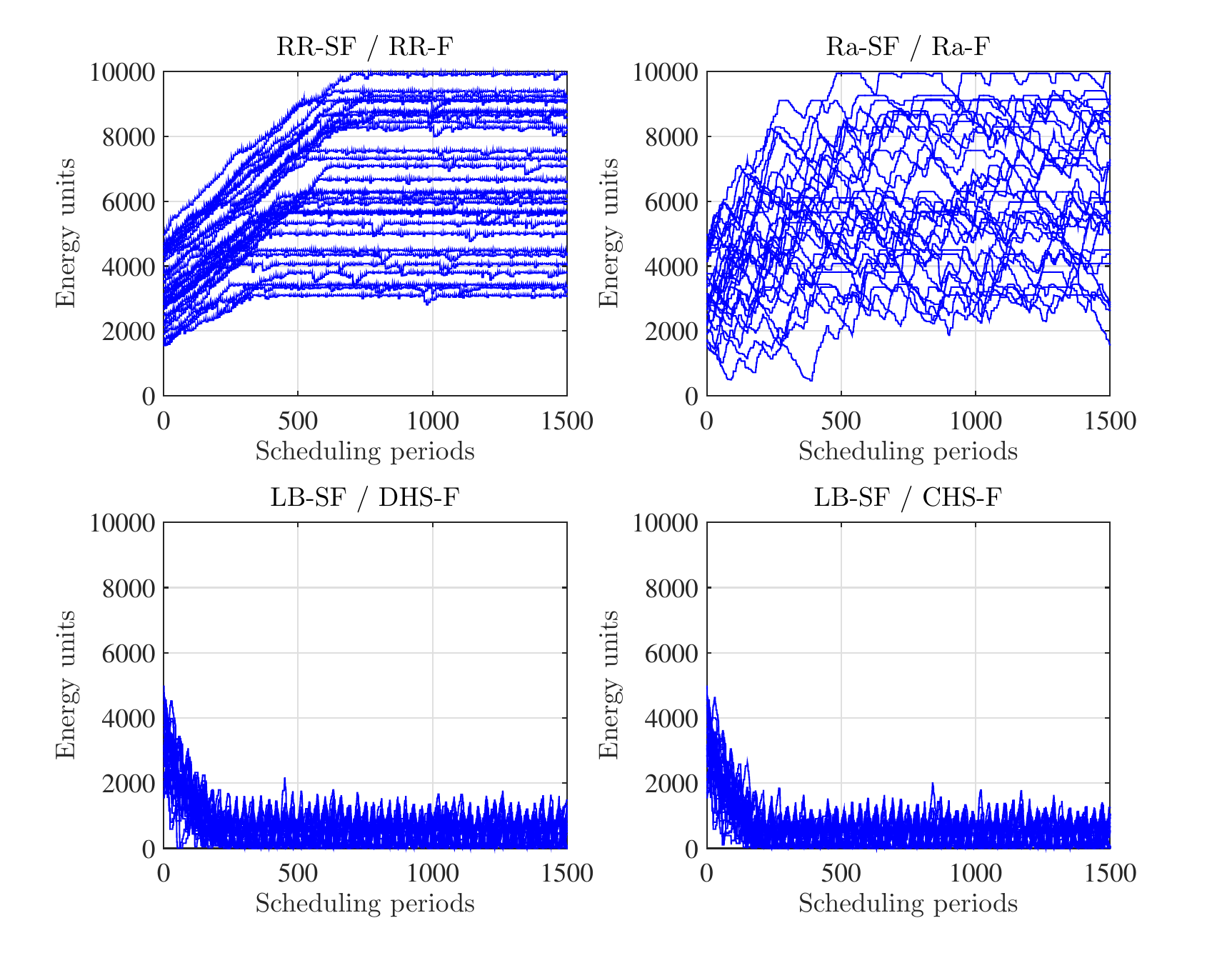}
	\caption{Time evolution of the battery levels of all users in the system for the different approaches.}
	\label{fig_sim1}
\end{figure}

Fig. \ref{fig_sim1} depicts the evolution of the battery levels of all users in the system. We can observe that for the round robin scheme, users reach their maximum battery capacity. This is because the data rates achieved are low, and thus, users use little energy for decoding. Then, in the top-right figure, we have the case where random scheduling is considered. In this case, we see how the battery evolutions of all users evolve randomly because new users are scheduled in a random fashion in each frame. Due to the battery overflows that some users experience and the randomness in the selection, this approach, as will happen with the round robin scheme, will not be very efficient in terms of aggregate throughput. The last two figures depict the battery evolution of the two proposed schemes. We observe that in both cases the battery levels of the users are substantially lower than the ones observed in previous schemes. This reduction in battery levels is related with the large throughput achieved by the users, as will be apparent later. It is difficult to assess from these figures which proposed scheme provides better performance.

\begin{figure}
	\centering
	\textbf{Time evolution of the average sum rate}\\
	\includegraphics[width=0.6\textwidth]{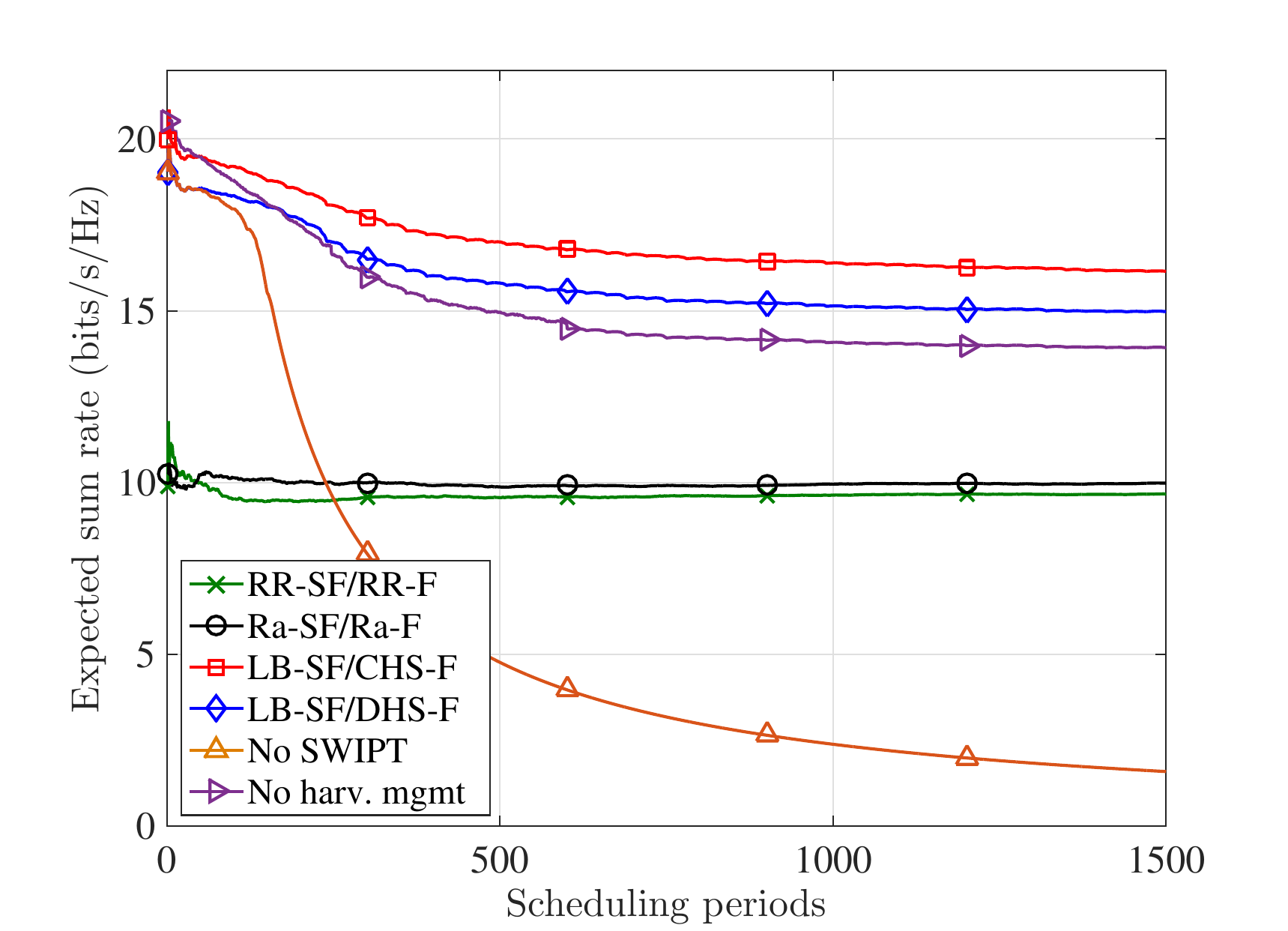}
	\caption{Time evolution of the average sum rate of the system for the different approaches.}
	\label{fig_sim2}
\end{figure}

Fig. \ref{fig_sim2} presents the average sum rate of the system (computed as $SR(\tau) = \frac{1}{\tau} \sum_{t=1}^\tau \sum_{i\in\mathcal{U}_I} R_i(t)$). This metric is an estimation of the expected throughput of the system. From the figure, we see that the sum rate of the round robin and random schemes provides a stable average throughput over time but the magnitude of the throughput is not so high. Then, we see how the proposed schemes notably outperform the previous benchmarking strategies. The simpler approach, DHS, performs similar to the more complex strategy, CHS. We also plot, as benchmarks, two cases. The first one, called 'no harvesting management', refers to the case in which the harvesting users are selected jointly with data users following the CHS approach, but their harvesting constraints are set to zero, $Q_j = 0,\, \forall j$, that is, harvesting users collect energy without imposing a constraint. In this case, the rate achieved is higher at the beginning, but the energy collected by the users is lower, having an impact on the performance as time goes on. The second case considers that no power transfer (no SWIPT) is available, and users therefore cannot recharge their batteries. In this case, the users run out of battery, and the expected sum rate therefore tends to zero. 

\begin{figure}
	\centering
	\textbf{CDF of the individual data rates}\\
	\includegraphics[width=0.6\textwidth]{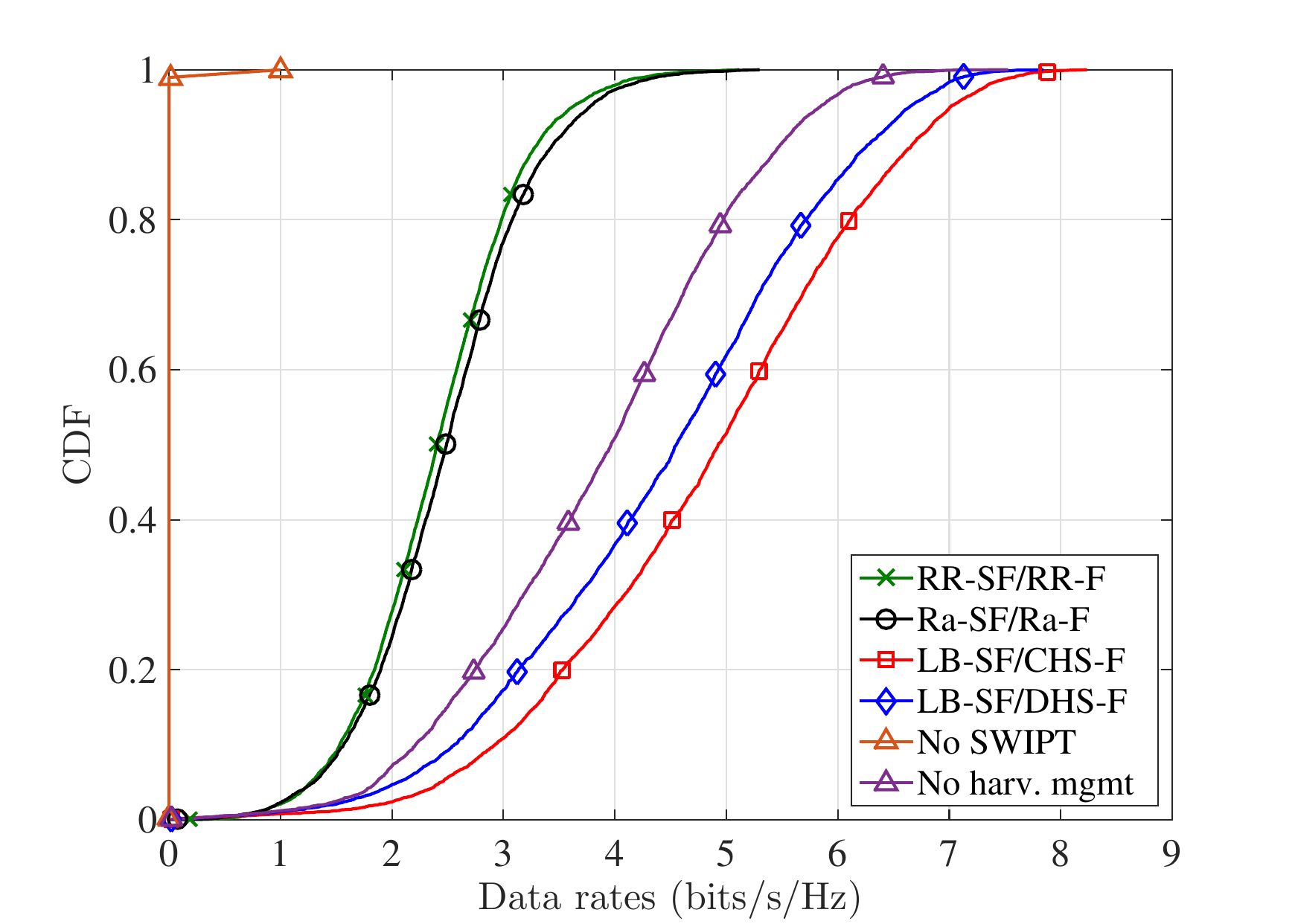}
	\caption{CDF of the individual data rates of all the users in the system for the different approaches.}
	\label{fig_sim3}
\end{figure}

Fig. \ref{fig_sim3} shows the cumulative distribution function (CDF) of the individual data rates of the users in the system. The CDF of the no SWIPT case has a particular shape due to the fact that many users obtain zero data rate as they run out of battery. In this figure, we clearly see the benefits of the proposed user selection schemes compared to the other approaches, such as low data rate percentiles and high data rate percentiles being much better for the proposed strategies. 

\begin{figure}
	\centering
	\textbf{Time evolution of the average harvested power}\\
	\includegraphics[width=0.6\textwidth]{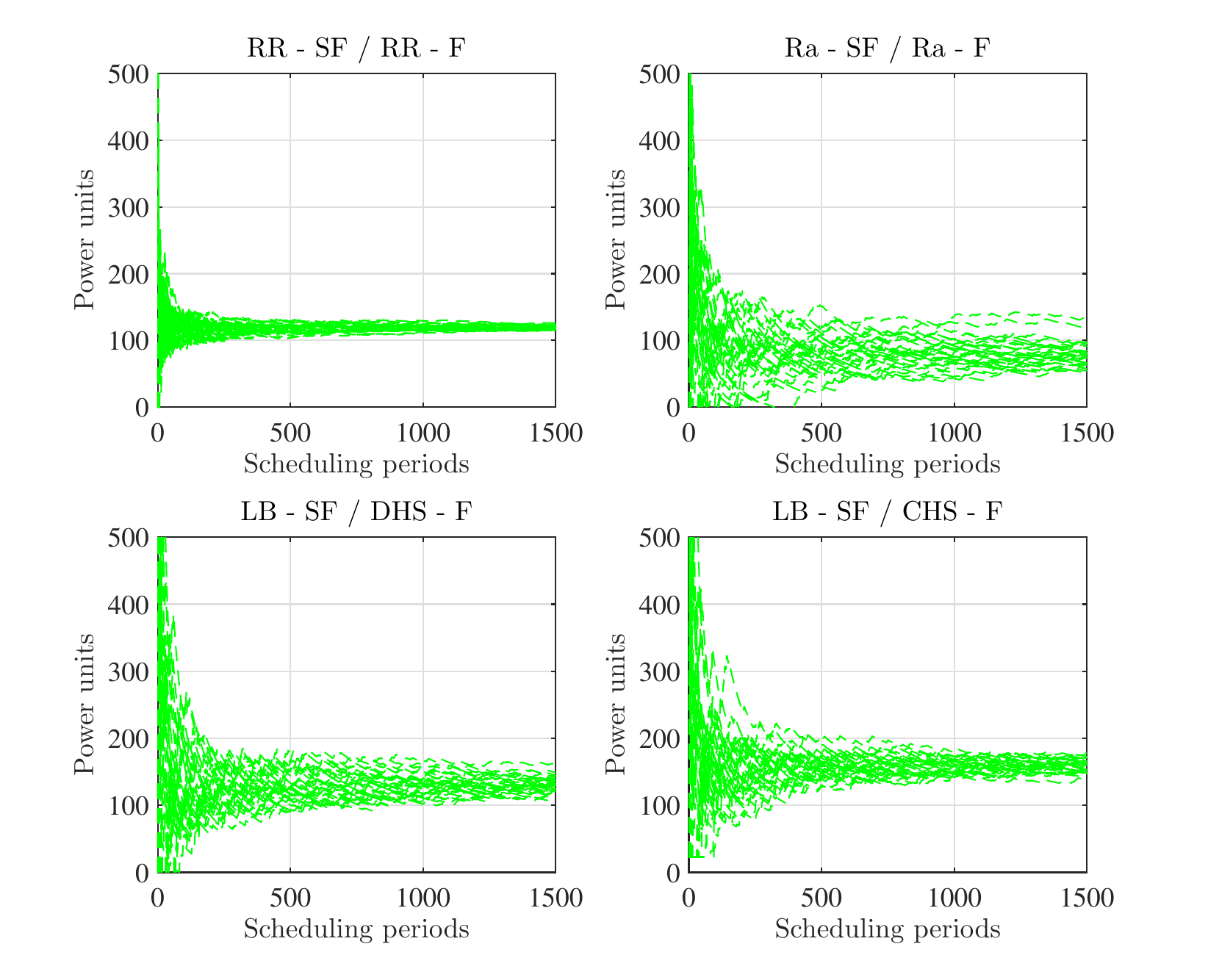}
	\caption{Time evolution of the average harvested power of all the users in the system for the different approaches (in power units).}
	\label{fig_sim4}
\end{figure}

Finally, in Fig. \ref{fig_sim4}, we depict the average evolution of the harvested power. It is interesting to note how all users tend to converge to a certain point (or the vicinity of a point). This is due to the fact that if a user is receiving much power, then its battery will increase, which will make the user more eligible to receive data, making the harvesting decrease, whereas if a user has low energy in its battery, then it is directly selected to be included in set $\mathcal{U}^S_E$. We observe that the more complex approach, CHS, is able to provide the users with larger harvested power compared to the less complex approach, DHS. 

\subsection{System Performance Simulations}
\begin{figure}
	\centering
	\textbf{Expected system sum rate vs distance to the BS}\\
	\includegraphics[width=0.6\textwidth]{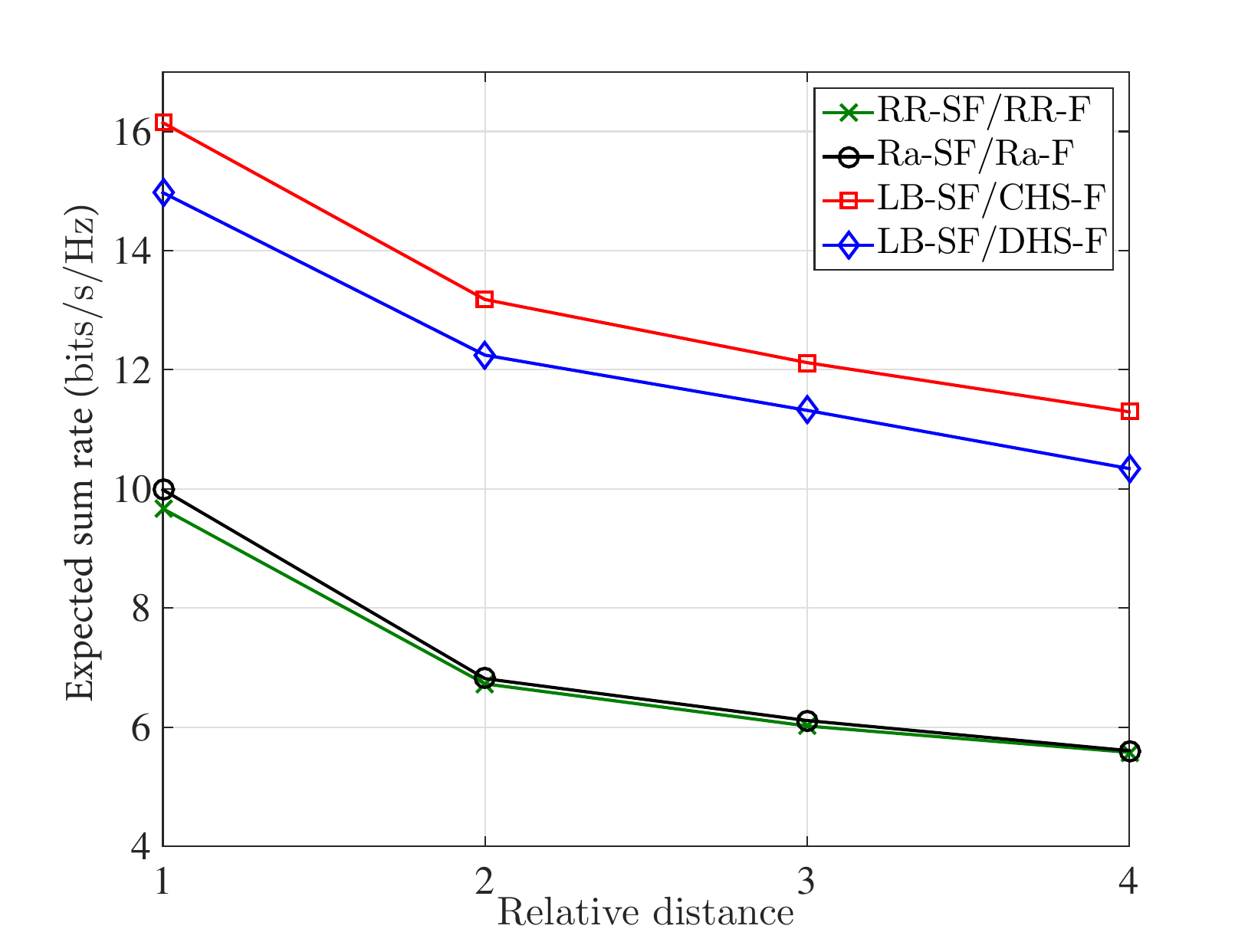}
	\caption{Expected system sum rate as a function of the distance to the BS.}
	\label{fig_sim5}
\end{figure}

\begin{figure}
	\centering
	\textbf{Sum of expected harvested powers by all users vs distance to the BS}\\
	\includegraphics[width=0.6\textwidth]{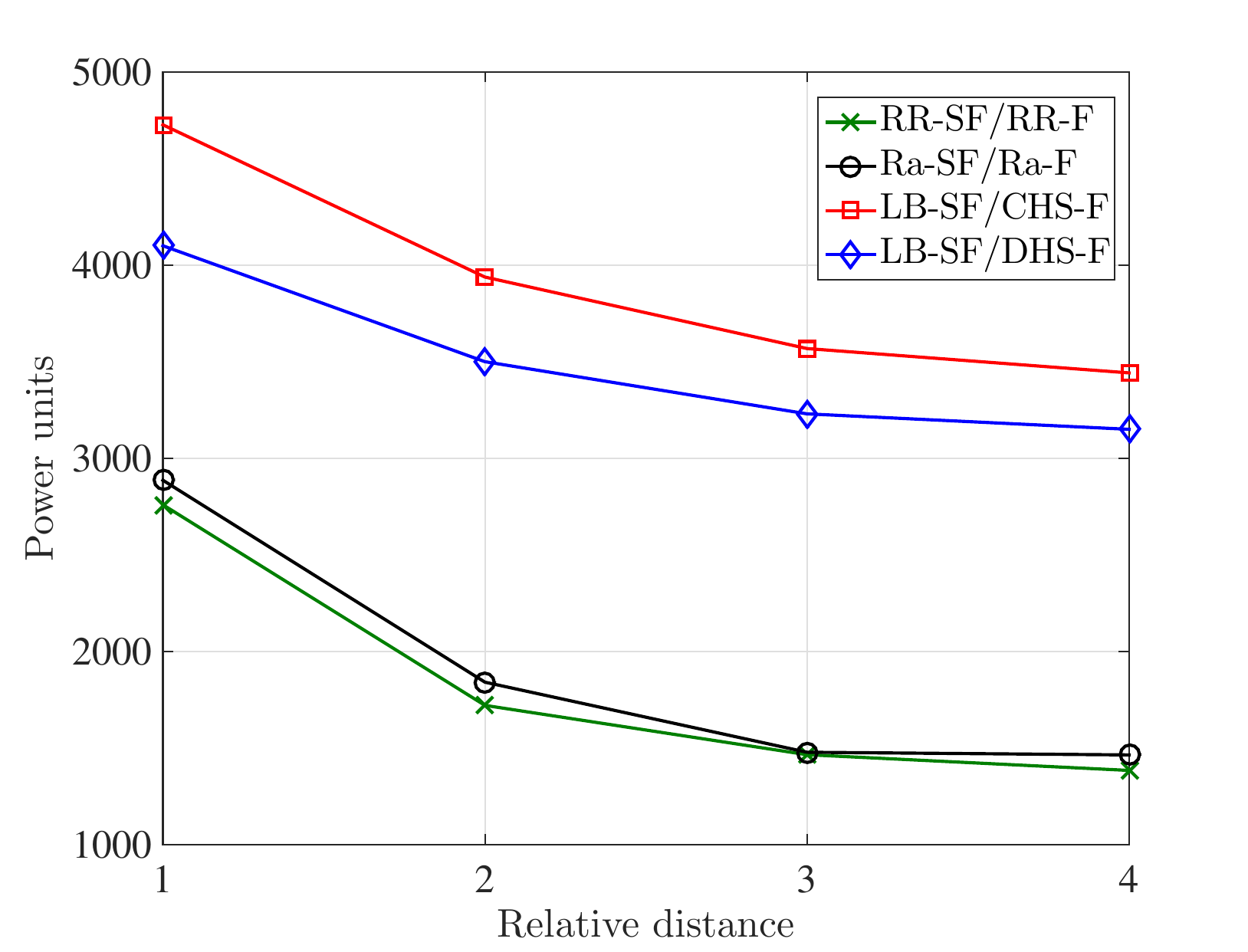}
	\caption{Sum of the expected harvested powers by all users (in power units) as a function of the distance to the BS.}
	\label{fig_sim6}
\end{figure}

In the next figures, we will show the performance of the system obtained once the algorithms have converged (i.e., after 1500 frames). The first two figures, Fig. \ref{fig_sim5} and Fig. \ref{fig_sim6}, show the system performance, considering that half of the users are at a relative distance to the BS greater than for the other half of the users. In particular, Fig. \ref{fig_sim5} presents the sum of the expected sum rate for the four schemes for four different relative distances. As expected, the sum rate decreases as the distance to the BS increases. On the other hand, Fig. \ref{fig_sim6} shows the sum of the expected harvested power as a function of the relative distance. We see that if half of the users are four times farther away from the BS, the loss in harvested power is from $25\%$ to $50\%$, and the relative loss is lower for the proposed schemes. 

\begin{figure}
	\centering
	\textbf{Expected system sum rate vs relative size of the harvesting user group}\\
	\includegraphics[width=0.6\textwidth]{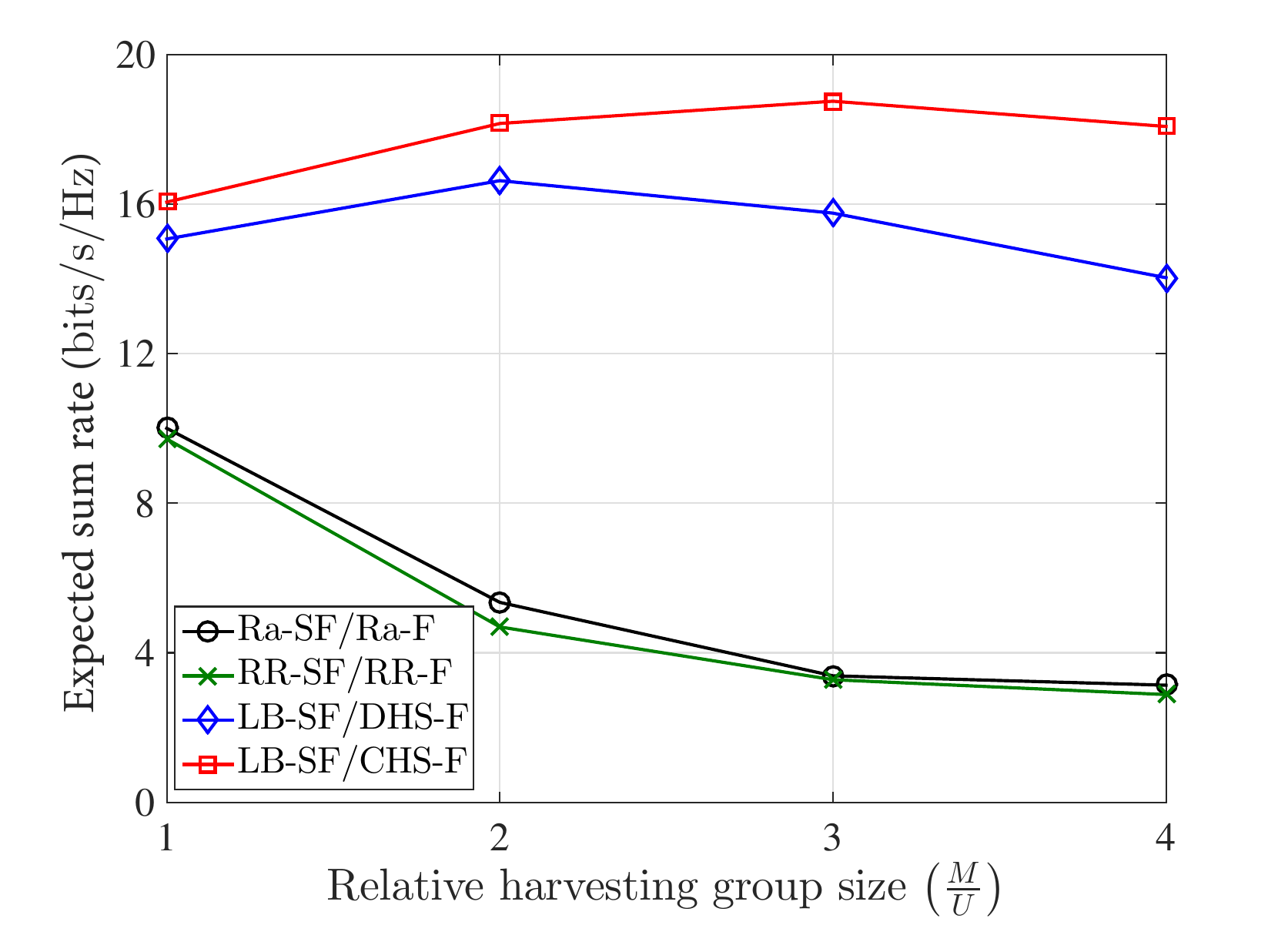}
	\caption{Expected system sum rate as a function of the relative size of the harvesting user group.}
	\label{fig_sim7}
\end{figure}

\begin{figure}
	\centering
	\textbf{Sum of expected harvested powers by all users vs relative size of the harvesting user group}\\
	\includegraphics[width=0.6\textwidth]{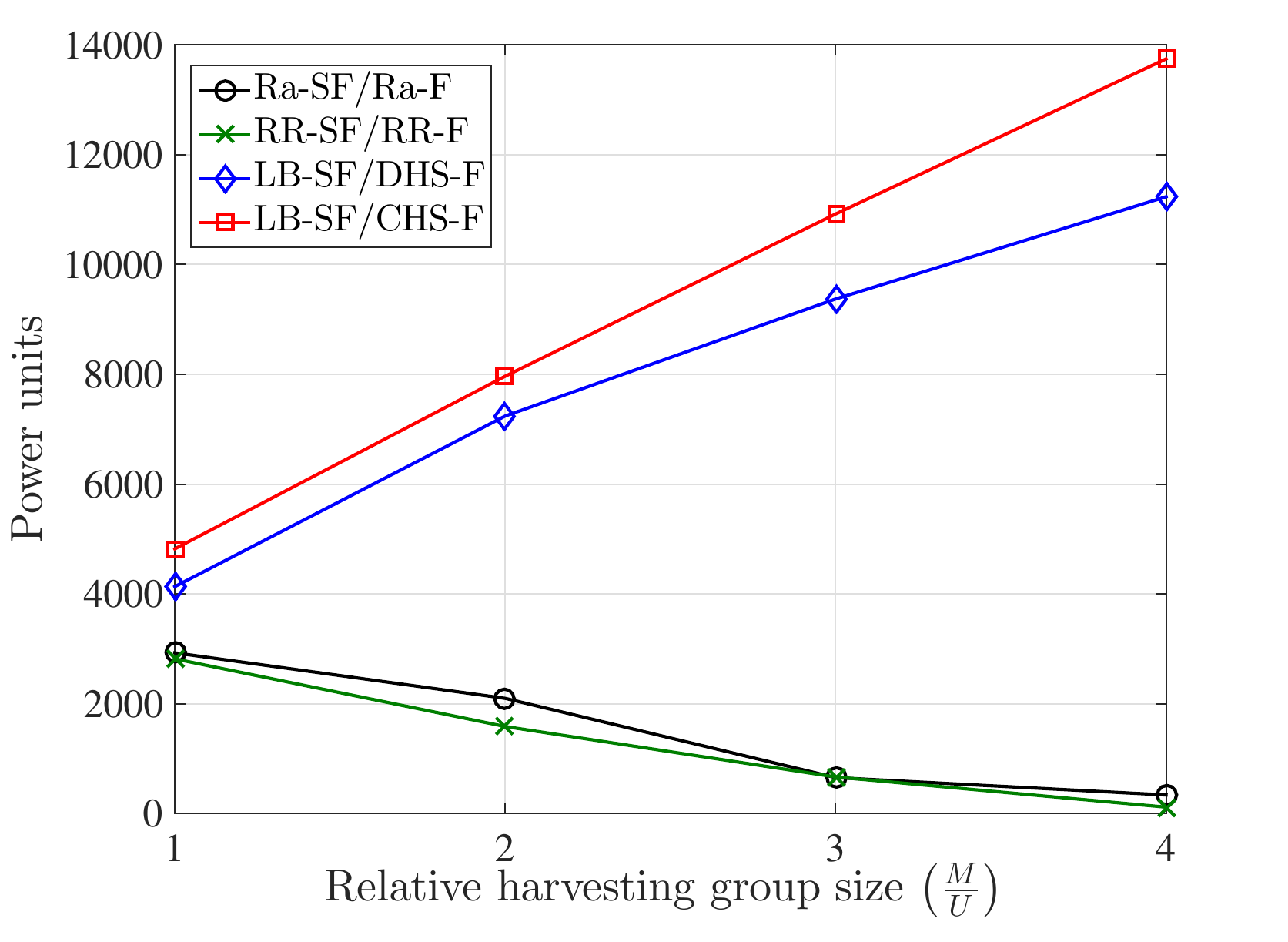}
	\caption{Sum of the expected harvested powers by all users (in power units) as a function of the relative size of the harvesting user group.}
	\label{fig_sim8}
\end{figure}

The last two figures, Fig. \ref{fig_sim7} and Fig. \ref{fig_sim8}, show the performance of the system when the size of the harvesting group increases in relative terms when compared to the size of the information group, i.e., when $\frac{M}{U}$ increases. This phenomenon is interesting to evaluate since the harvesting users appear in the constraints and they negatively affect the aggregated sum rate (see tradeoff in Section \ref{trade-off}). However, if many users are introduced in the harvesting set, then their batteries will recharge faster, and they will be able to receive higher data rates. This is the compromise that is analyzed in the figures. First, in Fig. \ref{fig_sim7}, we see the expected aggregated sum rate. As we see, for the two benchmarking approaches, Ra and RR, the sum rate decreases as $\frac{M}{U}$ increases. This is because the harvesting users are selected without considering the impact that they have on the objective function, and therefore, if more harvesting users are considered in the optimization problem, a lower sum rate will be achieved. In those cases, the optimization problem turns out to be infeasible many times, and therefore, the energy collected by all users also decreases--see Fig. \ref{fig_sim8}. On the other hand, the aggregated sum rate increases a bit for $\frac{M}{U} = 2$ for the proposed strategies. This is due to the fact that harvesting users are selected very efficiently, and thus, the constraints associated to them are not active, i.e., they do not affect the optimum value of the objective function. Additionally, as more users are able to recharge their batteries (see Fig. \ref{fig_sim8}), they can decode higher rates in future frames. Nonetheless, from a given size $\frac{M}{U}$ on, the system sum rate starts to decrease as the harvesting constraints become active, although the problem is always feasible, and users recharge their batteries, as is indirectly depicted in Fig. \ref{fig_sim8}. 

\subsection{Computational Complexity}
\begin{figure}
	\centering
	\textbf{Relative computational complexities}\\
	\includegraphics[width=0.6\textwidth]{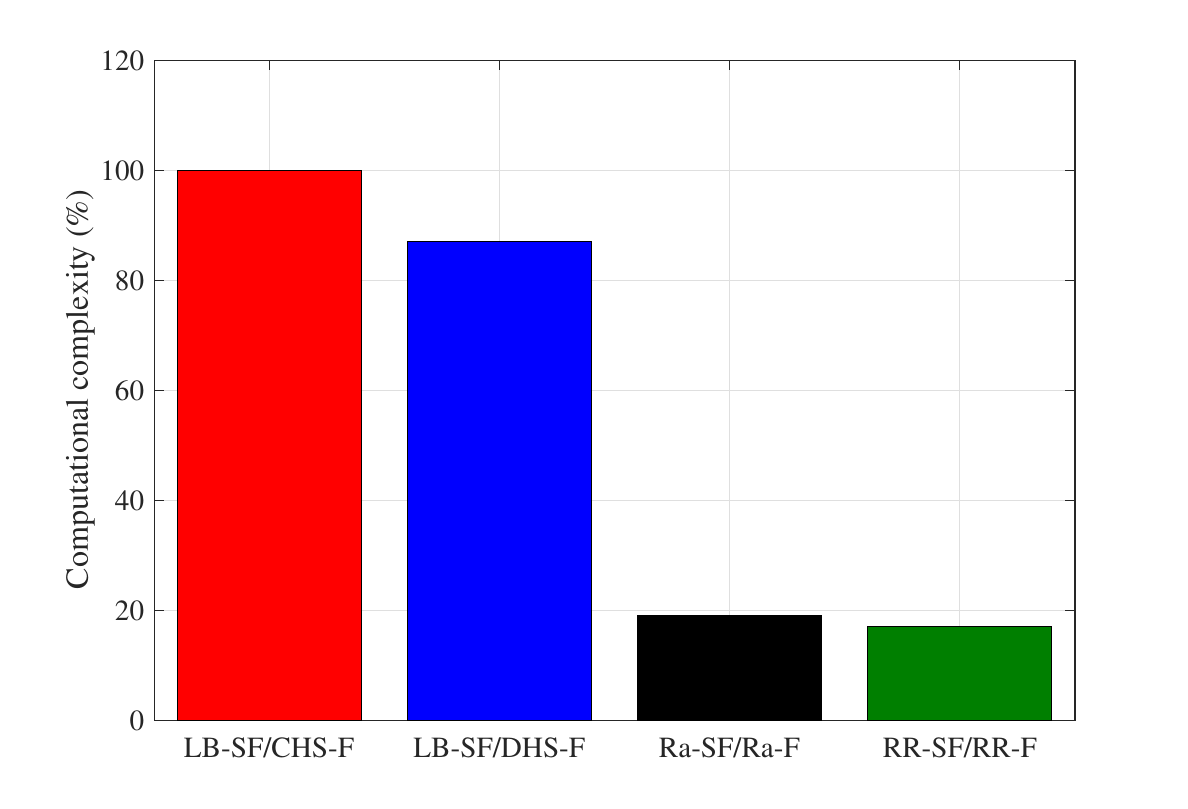}
	\caption{Average computational complexities of each algorithm needed for convergence at each scheduling period relative to algorithm LB-SF/CHS-F.}
	\label{fig_comput_complex}
\end{figure}

An analytic evaluation of the computational complexity of the proposed techniques for each scheduling period is extremely difficult since these algorithms are iterative, each iteration involves the numerical solution of an optimization problem, there are discrete variables related to the grouping of users, and the solution and convergence times depend on the concrete channels associated to the users in the scenario. Because of this, we have performed a numerical evaluation of the computational complexity of the different algorithms by performing many simulations over random channels and averaging the convergence times at each scheduling period obtained in the simulator. Fig. \ref{fig_comput_complex} shows a set of bars comparing the complexities needed for convergence of the different algorithms that require grouping, that is, RR-SF/RR-F, Ra-SF/Ra-F, LB-SF/CHS-F, and LB-SF/DHS-F. The highest bar corresponds to the algorithm requiring the highest computational complexity, which is LB-SF/CHS-F and has been labeled as the 100\% reference. The other bars show the complexities associated to the other algorithms, taking as relative reference, the complexity of LB-SF/CHS-F.

\section{Conclusions}
\label{sec_con}
This paper has studied the performance of a proposed scheduling algorithm in a multiuser MIMO broadcasting system, where wireless power transfer from BS has been considered a potential technique for energy harvesting taken from radio signals. We derived the particular structure of the optimal transmit covariance matrices and particularized the scenario where only information or harvesting users were present in the system and where both types of users coexist in the system. If only harvesting users were considered, the problem was reformulated as a feasibility problem, and we provided some proposals to be applied in case that the original problem was infeasible. Then, we addressed the multidimensional tradeoff between the sum rate and the harvesting constraints in the general case. We showed that energy beamforming was optimal in the case that the power harvested by one particular user was to be maximized. Finally, we presented some user grouping techniques that allow for the BS to select the users better suited for information and those for battery replenishment in each particular frame for the case, where both types of users are present in the system. We proposed two different scheduling techniques based on a different level of computational complexity. In the first approach, we selected the information and the harvesting users separately. In the second approach, the selection of both types of users was performed jointly. The simulation results show that the aggregated throughput can be considerably improved if the proposed grouping strategy is implemented when the results are compared with those of traditional scheduling approaches.

\appendix{}

\subsection{\textbf{Appendix}}
\label{app2}
The Lagrangian of problem \eqref{op:wet5} is
\begin{eqnarray}
\mathcal{L}(\{\tilde{\textbf{S}}_i\};\boldsymbol\lambda,\mu) &=&-\sum_{i\in\mathcal{U}_I}\omega_i \log \det \left(\textbf{I} + \hat{\textbf{H}}_i\tilde{\textbf{S}}_i\hat{\textbf{H}}_i^H\right)   \\
& &+ \sum_{j\in\mathcal{U}_E} \lambda_j \left(Q_j - \sum_{i\in\mathcal{U}_I}\Tr(\hat{\textbf{H}}_{ji}\tilde{\textbf{S}}_i\hat{\textbf{H}}_{ji}^H) \right)+\mu \left(\sum_{i\in\mathcal{U}_I}\Tr(\tilde{\textbf{S}}_i) - P_T\right)\nonumber
\end{eqnarray}
where we have omitted constraint $C3$. The previous Lagrangian can be manipulated and transformed into
\begin{equation}
\mathcal{L}(\{\tilde{\textbf{S}}_i\};\boldsymbol\lambda,\mu) =- \sum_{i\in\mathcal{U}_I} \omega_i\log \det \left(\textbf{I} + \hat{\textbf{H}}_i\tilde{\textbf{S}}_i\hat{\textbf{H}}_i^H\right)   + \sum_{i\in\mathcal{U}_I} \Tr\left(\textbf{A}_i \tilde{\textbf{S}}_i\right) + G,\label{lag}
\end{equation}
where $G = \sum_{j\in\mathcal{U}_E} \lambda_j Q_j - \mu P_T$, and $\textbf{A}_i = \mu \textbf{I} - \sum_{j\in\mathcal{U}_E} \lambda_j \hat{\textbf{H}}_{ji}^H\hat{\textbf{H}}_{ji}$.
The dual function of problem \eqref{op:wet5} is defined as $g(\boldsymbol\lambda, \mu) = \min_{\tilde{\textbf{S}}_i\succeq 0} \mathcal{L}(\{\tilde{\textbf{S}}_i\};\boldsymbol\lambda,\mu)$.

\begin{proposition}
	To have a bounded solution of the dual function $g(\boldsymbol\lambda, \mu)$, matrix $\textbf{A}_i$ must be $\textbf{A}_i \succ 0\,\, \forall i$; otherwise, $g(\boldsymbol\lambda, \mu)$ is unbounded below, i.e., $g(\boldsymbol\lambda,\mu) = -\infty$.
\end{proposition}
\begin{proof}
	See Appendix \ref{app3}.
\end{proof}
Due to the fact that matrices $\{\textbf{A}_i\}$ are positive definite, we can assure that they can be decomposed as $\textbf{A}_i = \textbf{A}_i^{1/2}\textbf{A}_i^{1/2}$ and that they always have inverses. Thus, by calling $\hat{\textbf{S}}_i = \textbf{A}_i^{1/2}\tilde{\textbf{S}}_i\textbf{A}_i^{1/2}$, the dual function can be expressed as
\begin{equation}
g(\boldsymbol\lambda,\mu) = \min_{\hat{\textbf{S}}_i\succeq 0} \bigg\{-\sum_{i\in\mathcal{U}_I} \omega_i\log \det \left(\textbf{I} + \hat{\textbf{H}}_i\textbf{A}_i^{-1/2}\hat{\textbf{S}}_i\textbf{A}_i^{-1/2}\hat{\textbf{H}}_i^H\right) + \sum_{i\in\mathcal{U}_I} \Tr\left(\hat{\textbf{S}}_i\right) + G\bigg\}.
\label{dual_fun}
\end{equation}
The dual function in \eqref{dual_fun} can be recognized to be equivalent to the dual function of the classical maximization of the sum rate with a power constraint, where the optimum covariance matrix $\hat{\textbf{S}}_i$ diagonalizes the equivalent channel $\hat{\textbf{H}}_i\textbf{A}_i^{-1/2}$ \cite{zhang:10a}, i.e., $\hat{\textbf{S}}_i = \hat{\textbf{V}}_i \hat{\textbf{D}}_i\hat{\textbf{V}}_i^H$, where $\hat{\textbf{D}}_i$ is the power allocation matrix, and its components are computed following the water-filling policy \cite{cover_book}. Finally, it is straightforward to show that the precoder $\textbf{B}_i$ matrix with dimensions $n_T\times n_{S_i}$ corresponding to such covariance matrix is
\begin{equation}
\textbf{B}_i^\star = \tilde{\textbf{V}}_i^{(0)}\textbf{A}_i^{-1/2} \hat{\textbf{V}}_i \hat{\textbf{D}}_i^{1/2}.
\end{equation}

\subsection{\textbf{Appendix}}
\label{app3}
Let the eigen-decomposition of $\textbf{A}_i$ be $\bar{\textbf{U}}_i\bar{\mathbf{\Gamma}}_i\bar{\textbf{U}}_i^H$, where $\bar{\mathbf{\Gamma}}_i$ contains the eigenvalues in decreasing order w.l.o.g. Then, the second term of the Lagrangian in \eqref{lag} is $\sum_{i\in\mathcal{U}_I}\Tr\left(\bar{\mathbf{\Gamma}}_i\bar{\textbf{U}}_i^H\tilde{\textbf{S}}_i\bar{\textbf{U}}_i\right)$. Now, calling $\bar{\textbf{S}}_i = \bar{\textbf{U}}_i^H\tilde{\textbf{S}}_i\bar{\textbf{U}}_i$, $(\bar{\textbf{S}}_i \succeq 0 \Longleftrightarrow \tilde{\textbf{S}}_i \succeq 0)$, and $\hat{\bar{\textbf{H}}}_i = \hat{\textbf{H}}_i\bar{\textbf{U}}_i$, we have $g(\boldsymbol\lambda,\mu) = \min_{\bar{\textbf{S}}_i\succeq 0} -\sum_{i\in\mathcal{U}_I} \omega_i\log \det \left(\textbf{I} + \hat{\bar{\textbf{H}}}_i\bar{\textbf{S}}_i\hat{\bar{\textbf{H}}}_i^H\right) + \sum_{i\in\mathcal{U}_I} \Tr\left(\bar{\mathbf{\Gamma}}_i \bar{\textbf{S}}_i\right) + G$. Let us take the particular structure for the covariance matrix $\bar{\textbf{S}}_i$ as being diagonal, with all the elements equal to 0, except the last one, which is equal to $P$, i.e., $\bar{\textbf{S}}_i = \textrm{diag}(0, \dots , P)$. Then, denoting $L_i = n_T- n_R + n_{R_i}$, the first term of the dual function becomes $-\sum_{i\in\mathcal{U}_I}\omega_i \log \left(1 + P \|[\hat{\bar{\textbf{H}}}_i]_{:,L_i}\|^2\right)$, where $[\hat{\bar{\textbf{H}}}_i]_{:,L_i}$ denotes the $L_i$-th column of $\hat{\bar{\textbf{H}}}_i$. Since matrix $\hat{\bar{\textbf{H}}}_i$ is formed by unitary rotations of a random matrix with i.i.d. entries, we can assure with probability equal to 1 that $\|[\hat{\bar{\textbf{H}}}_i]_{:,L_i}\|^2 \neq 0$. As a conclusion, the first term of the Lagrangian is negative and decreases without bound as $P$ increases. Let us have a look at the second term. If matrix $\textbf{A}_i$ is not positive definite, i.e., if the lowest element (and, thus, the last component) in the diagonal of $\bar{\mathbf{\Gamma}}_i$ is not positive, then the second term of the Lagrangian either is negative and decreases without bound as $P\rightarrow \infty$ or is zero. In both cases, and taking into account the behavior of the first term of the Lagrangian as $P$ tends to infinity, it is concluded that the dual function is equal to $-\infty$. Thus, the only possible solution so that $g(\boldsymbol\lambda, \mu) \neq - \infty$ is that $\bar{\mathbf{\Gamma}}_i$ has diagonal elements that are all strictly positive and, thus, $\textbf{A}_i \succ 0$.

\subsection{\textbf{Appendix}}
\label{part_cases}
\subsubsection{System with only information users}
Let us consider first the broadcast scenario with only users to be served with information and no energy harvesting users, i.e., $\mathcal{U}_E = \emptyset$. In this case, problem \eqref{op:wet1} can be expressed as
\begin{alignat}{3}
\mathop{\text{maximize}}_{\{R_i, \,\textbf{S}_i\}_{\forall i\in\mathcal{U}_I}}& \quad \sum_{i\in\mathcal{U}_I} \omega_iR_i \label{op:wet2} \\
\textrm{subject to}
&  \quad C2\dots C6\text{ of problem } \eqref{op:wet_g}.\nonumber
\end{alignat}
Without going into too much detail, let us say that the optimal solution to the above problem was presented in \cite{rub:14a} and is omitted due to space limitations.

\subsubsection{System with only harvesting users}
Let us now consider the case where there are only users who want to harvest energy, i.e., $\mathcal{U}_I = \emptyset$. In this case, since there is no objective function, the optimization problem becomes a feasibility problem \cite{boyd} that can be expressed as\footnote{The case of not having information users is special as the harvesting users cannot take advantage of the spurious signals intended for the information users to recharge their batteries. Only in this case, we allow for the base station to send a specific signal to the harvesting users and those whose covariance matrix is defined by \textbf{S}.}
\begin{alignat}{2}
\mathop{\text{find}}& \quad \textbf{S} \label{op:wet3} \\
\textrm{subject to}
&  \quad C1, C2, C6\text{ of problem } \eqref{op:wet_g}. \nonumber
\end{alignat}
Note that constraints $C3$, $C4$, and $C5$ from problem \eqref{op:wet1} have no effect since the set $\mathcal{U}_I$ is empty. Notice also that, without loss of optimality, we have changed the optimization variable from a set of precoding matrices $\{\textbf{S}_i\}$ to a single precoder matrix $\textbf{S}$. In the following, we will present a necessary condition for feasibility of \eqref{op:wet3}. 

\begin{proposition}[\cite{bhatia_book}] \label{prop_eig}
	Let $\lambda_{\max}(\textbf{X}) \ge \lambda_2(\textbf{X}) \ge \dots \ge \lambda_{\min}(\textbf{X})$ be the eigenvalues of the positive semidefinite matrix $\textbf{X}$. Then, for any two semidefinite positive matrices, $\textbf{A}$ and $\textbf{B}$, we have
	\begin{eqnarray}
	\lambda_j(\textbf{AB}) \le \lambda_{\max}(\textbf{B})\lambda_j(\textbf{A}) \text{ and } \lambda_j(\textbf{BA}) \le \lambda_{\max}(\textbf{B})\lambda_j(\textbf{A}),\quad \forall j,\\
	\lambda_j(\textbf{AB}) \ge \lambda_{\min}(\textbf{B})\lambda_j(\textbf{A}) \text{ and } \lambda_j(\textbf{BA}) \ge \lambda_{\min}(\textbf{B})\lambda_j(\textbf{A}),\quad \forall j. 
	\end{eqnarray}
\end{proposition}

Note that the previous lemma can be generalized as $\Tr\left(\textbf{A}\textbf{B}\right) \le \lambda_{\max}(\textbf{B})\Tr(\textbf{A})$ since $\Tr(\textbf{A}) = \sum_j \lambda_j(\textbf{A})$. The inequality is attained when $\textbf{A}$ has rank 1 and is built with the eigenvector associated with the maximum eigenvalue of $\textbf{B}$, ($\textbf{e}_{\max}(\textbf{B})$), i.e., $\textbf{A} = k\,\textbf{e}_{\max}(\textbf{B})\textbf{e}_{\max}(\textbf{B})^H$.

\begin{proposition}\label{pro_nec}
	Let $\textbf{H}_j^H\textbf{H}_j= \textbf{V}_{H,j}\mathbf{\Sigma}_{H,j}\textbf{V}_{H,j}^H$ be the reduced eigenvalue decomposition of matrix $\textbf{H}_j^H\textbf{H}_j$ with $\mathbf{\Sigma}_{H,j} = \textrm{diag}(\sigma_{1,j}, \dots , \sigma_{n_{R_j},j})$ and $\sigma_{1,j} \ge \sigma_{2,j} \ge \dots \ge \sigma_{n_{R_j},j} > 0$. Then, a necessary condition for the feasibility of problem \eqref{op:wet3} is $(P_{\max} - P^{tx}_c)\sigma_{1,j} - Q_j \ge 0$, $\forall j$.
\end{proposition}
\begin{proof}
	Just as we considered before, if the problem is feasible, at least one solution fulfills $\Tr(\textbf{S}) = P_{\max} - P^{tx}_c$, and the maximum value that $\Tr(\textbf{H}_j\textbf{S}\textbf{H}_j^H)$ can take, based on Proposition \ref{prop_eig}, is $(P_{\max} - P^{tx}_c)\sigma_{1,j}$ $(\Tr(\textbf{AB}) \le \lambda_{\max}(\textbf{B})\Tr(\textbf{A}))$ with $\textbf{S} = (P_{\max} - P^{tx}_c)\textbf{v}_{n_{R_j},j}\textbf{v}_{n_{R_j},j}^H$, where $\textbf{v}_{n_{R_j},j}$ is the eigenvector associated with the maximum eigenvalue $\sigma_{1,j}$ of $\textbf{H}_j^H\textbf{H}_j$.
\end{proof}

Generally, as we are not able to provide a necessary and sufficient condition, we need to solve the following convex optimization problem to test the feasibility of problem \eqref{op:wet3}:
\begin{alignat}{2}
\mathop{\text{minimize}}_{\textbf{S},\bar{P}_{\max}}& \quad \bar{P}_{\max} \label{op:feas} \\
\textrm{subject to}
&  \quad C2:\Tr(\textbf{S}) + P^{tx}_c\le \bar{P}_{\max}\nonumber\\
&  \quad C1, C6\text{ of problem } \eqref{op:wet_g}. \nonumber
\end{alignat}
The above problem is categorized as a semidefinite optimization problem. There is no closed-form solution for the above problem, but the optimum solution can be obtained efficiently with the application of interior point methods \cite{boyd}. Let us denote the optimum solution of the problem above as $\bar{P}_{\max}^\star$. Now, it only remains to check whether $\bar{P}_{\max}^\star \le P_{\max}$ (which means that the problem is feasible) or $\bar{P}_{\max}^\star > P_{\max}$ (which implies infeasibility). If the problem is feasible, the optimum covariance matrix obtained in \eqref{op:feas} is the matrix that fulfills all the harvesting power constraints with the minimum transmitted power. If the problem is infeasible, one possible solution would be to reduce all the power harvesting constraints $\{Q_j\}$ such that constraints $C1$ become looser until the problem becomes feasible.

\section*{List of abbreviations}

\begin{tabular}{ll}
AWGN & Additive White Gaussian Noise\\
BD & Block-Diagonalization\\
BS & Base Station\\
CDF & Cumulative Density Function\\
CSI & Channel State Information\\
DP & Dynamic Programming\\
HPA & High Power Amplifier\\
MIMO & Multiple-Input Multiple-Output\\
MISO & Multiple-Input Single-Output\\
MUI & Multiuser Interference\\
PF & Proportional Fair\\
R-P & Rate-Power\\
RF & Radio Frequency \\
SINR & Signal to Interference plus Noise Ratio \\
SISO & Single-Input Single-Output\\
SNR & Signal to Noise Ratio \\
SR & Sum Rate\\
SVD & Singular Value Decomposition\\
SWIPT & Simultaneous Wireless Information and Power Transfer \\
w.l.o.g. & Without Loss of Generality
\end{tabular}

\section*{Methods/Experimental}
The aim of the work presented in this paper is to develop a dynamic grouping mechanism that decides which users should be scheduled to receive information and which users should be configured to harvest energy. The design also includes the derivation of the optimal transmission covariance matrices.

The design is based on a theoretical modeling of the scenario and the signal, the definition of a mathematical optimization problem, and the proposal of an algorithm to find a suboptimal solution to that problem that can be implemented.

Finally, numerical computer simulations have been carried out to evaluate the performance of the proposed strategy based on the mathematical modeling of the setup.

\section*{Availability of data and materials}
Not applicable.

\section*{Competing interests}
The authors declare that they have no competing interests.

\section*{Funding}
The research leading to these results has received funding through the grant 2017 SGR 578 (funded by the Catalan Government—Secretaria d’Universitats i Recerca, Departament d’Empresa i Coneixement, Generalitat de Catalunya, AGAUR) and project TEC2016-77148-C2-1-R (AEI/FEDER, UE): 5G\&B-RUNNER-UPC (funded by
the Agencia Estatal de Investigación, AEI, and Fondo Europeo de Desarrollo Regional, FEDER).

\section*{Authors' contributions}
JR and API put forward the idea and wrote the manuscript. JR carried out the experiments and simulations. JR and API contributed to the interpretation of the results and read and approved the final manuscript.

\section*{Authors' information}
Email addresses: \texttt{javier.rubio.lopez@upc.edu} (Javier Rubio), \texttt{antonio.pascual@upc.edu} (Antonio Pascual-Iserte).

\bibliographystyle{unsrt}
\bibliography{referencias}

\begin{thebibliography}{10}

\bibitem{rub:13d}
J.~Rubio and A.~Pascual-Iserte.
\newblock Simultaneous wireless information and power transfer in multiuser {MIMO} systems.
\newblock In {\em IEEE Global Communications Conference ({GLOBECOM})}, pages 2755--2760, Dec. 2013.

\bibitem{paradiso:05}
J.A. Paradiso and T.~Starner.
\newblock Energy scavenging for mobile wireless electronics.
\newblock {\em IEEE Computing Pervasive}, 4(1):18--27, Jan. 2005.

\bibitem{sudevalayam:11}
S.~Sudevalayam and P.~Kulkarni.
\newblock Energy harvesting sensor nodes: survey and implications.
\newblock {\em IEEE Communications Surveys \& Tutorials}, 13(3):443--461, Third Quarter 2011.

\bibitem{zhang:19}
Z.~Zhang, H.~Pang, A.~Georgiadis, and C.~Cecati.
\newblock Wireless power transfer — an overview.
\newblock {\em IEEE Trans. on Industrial Electronics}, 66(2):1044--1058, Febr. 2019.

\bibitem{chandrasekhar:08}
V.~Chandrasekhar, J.~Andrews, and A.~Gatherer.
\newblock Femtocell networks: a survey.
\newblock {\em IEEE Comm. Magazine}, 46:59--67, Sep. 2008.

\bibitem{lu:15}
X.~Lu et~al.
\newblock Wireless networks with {RF} energy harvesting: A contemporary survey.
\newblock {\em IEEE Communications Surveys and Tutorials}, 17(2):757--789, Secondquarter 2015.

\bibitem{varshney:08}
L.~R. Varshney.
\newblock Transporting information and energy simultaneously.
\newblock In {\em International Symposium on Information Theory}, Jul. 2008.

\bibitem{grover:10}
P.~Grover and A.~Sahai.
\newblock Shannon meets {T}esla: wireless information and power transfer.
\newblock In {\em International Symposium on Information Theory}, Jun. 2010.

\bibitem{zhang:13b}
R.~Zhang and C.~K. Ho.
\newblock {MIMO} broadcasting for simultaneous wireless information and power transfer.
\newblock {\em IEEE Trans. on Wireless Communications}, 12(5):1989--2001, May 2013.

\bibitem{park:13}
J.~Park and B.~Clerckx.
\newblock Joint wireless information and energy transfer in a two-user {MIMO} interference channel.
\newblock {\em IEEE Trans. on Wireless Comm.}, 12(8):4210--4221, Aug. 2013.

\bibitem{park:14}
J.~Park and B.~Clerckx.
\newblock Joint wireless information and energy transfer in a {K}-user {MIMO} interference channel.
\newblock {\em IEEE Trans. on Wireless Comm.}, 13(10):5781--5796, Oct. 2014.

\bibitem{zong:16}
Z.~Zong et~al.
\newblock Optimal transceiver design for {SWIPT} in {K}-user {MIMO} interference channels.
\newblock {\em IEEE Trans. on Wireless Comm.}, 15(1):430--445, Jan. 2016.

\bibitem{xu:14}
J.~Xu, L.~Liu, and R.~Zhang.
\newblock Multiuser {MISO} beamforming for simultaneous wireless information and power transfer.
\newblock {\em IEEE Trans. on Signal Processing}, 62(18):4798--4810, Sep. 2015.

\bibitem{shi:14}
Q.~Shi, L.~Liu, W.~Xu, and R.~Zhang.
\newblock Joint transmit beamforming and receive power splitting for {MISO} {SWIPT} systems.
\newblock {\em IEEE Trans. on Wireless Comm.}, 13(6):3269--3280, Jun. 2014.

\bibitem{zeng:15}
Y.~Zeng and R.~Zhang.
\newblock Optimized training design for wireless energy transfer.
\newblock {\em IEEE Trans. on Communication}, 63(2):536--550, Feb. 2015.

\bibitem{xu:16}
J.~Xu and R.~Zhang.
\newblock A general design framework for {MIMO} wireless energy transfer with limited feedback.
\newblock {\em IEEE Trans. on Signal Processing}, 64(10):2475--2488, Feb. 2016.

\bibitem{xiang:12}
Z.~Xiang and M.~Tao.
\newblock Robust beamforming for wireless information and power transmission.
\newblock {\em IEEE Wireless Communications Letters}, 1(4):372--375, Aug. 2012.

\bibitem{shepard:12}
C.~Shepard et~al.
\newblock Argos: Practical many-antenna base stations.
\newblock In {\em ACM International Conference on Mobile Computing and Networking}, 2012.

\bibitem{zhang:13a}
L.~Liu, R.~Zhang, and K.-C. Chua.
\newblock Wireless information transfer with opportunistic energy harvesting.
\newblock {\em IEEE Trans. on Wireless Communications}, 12(1):288--300, Jan. 2013.

\bibitem{morsi:15}
R.~Morsi, D.S. Michalopoulos, and R.~Schober.
\newblock Multiuser scheduling schemes for simultaneous wireless information and power transfer over fading channels.
\newblock {\em IEEE Trans. on Wireless Comm.}, 14(4):1967--1982, Apr. 2015.

\bibitem{ding:14}
Z.~Ding and H.V. Poor.
\newblock User scheduling in wireless information and power transfer networks.
\newblock In {\em IEEE International Conference on Communication Systems (ICCS)}, Nov. 2014.

\bibitem{shen:15}
W-L. Shen, K.C-J. Lin, M-S. Chen, and K.~Tan.
\newblock {SIEVE}: Scalable user grouping for large mu-mimo systems.
\newblock In {\em IEEE Conference on Computer Communications ({INFOCOM})}, pages 1975--1983, Apr. 2015.

\bibitem{dau:19}
L.~Dai, B.~Wang, M.~Peng, and S.~Chen.
\newblock Hybrid precoding-based millimeter-wave massive {MIMO}-{NOMA} with simultaneous wireless information and power transfer.
\newblock {\em IEEE Journal on Sel. Areas in Comm.}, 37(1):131--141, Jan. 2019.

\bibitem{goldsmith:03}
A.~Goldsmith et~al.
\newblock Capacity limits of {MIMO} channels.
\newblock {\em IEEE Journal on Sel. Areas in Comm.}, 21(5):684--702, Jun. 2003.

\bibitem{spencer:04}
Q.~H. Spencer et~al.
\newblock Zero-forcing methods for downlink spatial multiplexing in multiuser {MIMO} channels.
\newblock {\em IEEE Trans. on Signal Processing}, 52(2):461--471, Feb. 2004.

\bibitem{zhang:10a}
R.~Zhang, Y.-C. Liang, and S.~Cui.
\newblock Dynamic resource allocation in cognitive radio networks.
\newblock {\em IEEE Signal Processing Magazine}, 27(3):102--114, May 2010.

\bibitem{rub:16a}
J.~Rubio, A.~Pascual Iserte, D.~P. Palomar, and A.~Goldsmith.
\newblock Joint optimization of power and data transfer in multiuser {MIMO} systems.
\newblock {\em IEEE Trans. on Signal Processing}, 65(1):212--227, Jan. 2017.

\bibitem{bertrand:12}
B.~Devillers and D.~Gündüz.
\newblock A general framework for the optimization of energy harvesting communication systems with battery imperfections.
\newblock {\em Journal of Communications and Networks}, 14(2):130--139, Apr. 2012.

\bibitem{jensen:12}
A.~R. Jensen et~al.
\newblock {LTE} {UE} power consumption model: for system level energy and performance optimization.
\newblock In {\em IEEE Vehicular Technology Conference ({VTC} 2012 Fall)}, Sep. 2012.

\bibitem{auer:11}
G.~Auer et~al.
\newblock How much energy is needed to run a wireless network?
\newblock {\em IEEE Trans. on Wireless Communications}, 18(5):40--49, Oct. 2011.

\bibitem{grover:11}
P.~Grover, K.~Woyach, and A.~Sahai.
\newblock Towards a communication-theoretic understanding of system-level power consumption.
\newblock {\em IEEE Journal on Sel. Areas in Comm.}, 29(8):1744--1755, Sep. 2011.

\bibitem{ros:10}
P.~Rost and G.~Fettweis.
\newblock On the transmission-computation-energy tradeoff in wireless and fixed networks.
\newblock In {\em IEEE Global Communications Conference ({GLOBECOM}), workshop on Green Communications}, pages 1934--1939, Dec. 2010.

\bibitem{rub:14a}
J.~Rubio and A.~Pascual-Iserte.
\newblock Energy-aware broadcast multiuser-{MIMO} precoder design with imperfect channel and battery knowledge.
\newblock {\em IEEE Trans. on Wireless Communications}, 13(6):3137--3152, Jun. 2014.

\bibitem{lee:06}
J.~Lee and N.~Jindal.
\newblock Dirty paper coding vs linear precoding for {MIMO} broadcast channels.
\newblock In {\em Asilomar Conference on Signals, Systems and Computers}, pages 779--783, Oct. 2006.

\bibitem{ho:12}
C.~Ho and R.~Zhang.
\newblock Optimal energy allocation for wireless communications with energy harvesting constraints.
\newblock {\em IEEE Trans. on Signal Processing}, 60(9):4808--4818, Sep. 2012.

\bibitem{yang:12}
J.~Yang, O.~Ozel, and S.~Ulukus.
\newblock Broadcasting with an energy harvesting rechargeable transmitter.
\newblock {\em IEEE Trans. on Wireless Communications}, 11(2):571--583, Feb. 2012.

\bibitem{bertsekas_dp}
D.~P. Bertsekas.
\newblock {\em Dynamic programming and optimal control}.
\newblock Athena Scientific, thrid edition, 2005.

\bibitem{blasco:13}
P.~Blasco, D.~Gündüz, and M.~Dohler.
\newblock A learning theoretic approach to energy harvesting communication system optimization.
\newblock {\em IEEE Trans. on Wireless Communications}, 12(4):1872--1882, Apr. 2013.

\bibitem{lopez_ramos:14}
L.~M. Lopez-Ramos, A.~G. Marques, and J.~Ramos.
\newblock Jointly optimal sensing and resource allocation for multiuser interweave cognitive radios.
\newblock {\em IEEE Trans. on Wireless Communications}, 13(11):5854--5967, Nov. 2014.

\bibitem{boyd}
S.~Boyd and L.~Vandenbergue.
\newblock {\em Convex optimization}.
\newblock Cambridge, 2004.

\bibitem{bland:81}
R.~G. Bland, D.~Goldfarb, and M.~J. Todd.
\newblock The ellipsoid method: a survey.
\newblock {\em Operations Research}, 29(6):1039--1091, 1981.

\bibitem{bertsekas}
D.~P. Bertsekas.
\newblock {\em Nonlinear programming}.
\newblock Athena Scientific, second edition, 1999.

\bibitem{rub:15a}
J.~Rubio and A.~Pascual-Iserte.
\newblock Harvesting management in multiuser {MIMO} systems with simultaneous wireless information and power transfer.
\newblock In {\em IEEE Vehicular Technology Conference ({VTC}) Spring}, May 2015.

\bibitem{dimic:05}
G.~Dimic and N.~Sidiropoulos.
\newblock On downlink beamforming with greedy user selection: performance analysis and a simple new algorithm.
\newblock {\em IEEE Trans. on Signal Processing}, 53(10):3857--3868, Oct. 2005.

\bibitem{sigdel:09}
S.~Sigdel and W.~A. Krzymien.
\newblock Simplified fair scheduling and antenna selection algorithms for multiuser {MIMO} orthogonal space-division mutiplexing downlink.
\newblock {\em IEEE Trans. on Vehicular Technology}, 58(3):1329--1344, Mar. 2009.

\bibitem{jalali:00}
A.~Jalali, R.~Padovani, and R.~Pankaj.
\newblock Data throughput of {CDMA-HDR} a high efficiency-high data rate personal communication wireless system.
\newblock In {\em IEEE Vehicular Technology Conference ({VTC}) spring,}, May. 2000.

\bibitem{wang:07}
X.~Wang, G.~B. Giannakis, and A.~G. Marques.
\newblock A unified approach to {Q}o{S}-guaranteed scheduling for channel-adaptive wireless networks.
\newblock {\em Proceedings of IEEE}, 95(12):2410--2431, Dec. 2007.

\bibitem{liu:10}
Lingjia Liu, Young-Han Nam, and Jianzhong Zhang.
\newblock Proportional fair scheduling for multi-cell multi-user {MIMO} systems.
\newblock In {\em 44th Annual Conference on Information Sciences and Systems (CISS)}, Mar. 2010.

\bibitem{cover_book}
T.~M. Cover and J.~A. Thomas.
\newblock {\em Elements of information theory}.
\newblock Wiley, 2006.

\bibitem{bhatia_book}
R.~Bhatia.
\newblock {\em Perturbation Bounds for Matrix Eigenvalues}.
\newblock Longman Scientific \& Technical, 1987.

\bibitem{hua:16}
K.~Huang, C.~Zhong, and G.~Zhu.
\newblock Some new trends in wirelessly powered communications.
\newblock {\em IEEE Wireless Communicatgions}, 23(2):19--27, Apr. 2016.

\end{thebibliography}

\section*{Figures}

\end{document}